\documentclass[longauth]{aa} 
\usepackage{graphicx}
\usepackage{txfonts}
\usepackage{natbib}
\usepackage{placeins}
\usepackage[colorlinks=true,linkcolor=blue!80,citecolor=blue!80]{hyperref}
\usepackage{graphicx}
\usepackage{amssymb}
\usepackage{amsmath}
\usepackage{xcolor}
\usepackage{subcaption}
\usepackage{longtable}
\usepackage{statmath}
\usepackage{multirow} 
\usepackage{orcidlink}

\usepackage{siunitx}
\usepackage{threeparttable}

\usepackage{enumitem}
\newcommand{\um}{\SI{}{\micro\meter}}

\newcommand\tcr[1]{\textcolor{black}{#1}}

\def\arcmin{\ifmmode {^{\prime}}\else $^{\prime}$\fi}
\def\arcsec{\ifmmode {^{\prime\prime}}\else $^{\prime\prime}$\fi}
\newcommand\qth{$^{\rm th}$}
\newcommand{\av}{A$_{V}$}
\newcommand{\Ha}{H$\alpha$}
\newcommand{\Hb}{H$\beta$}
\newcommand{\sii}{[{S\sc{ii}}]}
\newcommand{\nii}{[{N\sc{ii}}]}
\newcommand{\oii}{[{O\sc{ii}}]}
\newcommand{\oiii}{[{O\sc{iii}}]}
\newcommand{\NH}{\nii/H$\alpha$}
\newcommand{\OTHb}{\oiii/\Hb}

\newcommand{\STHa}{\sii/\Ha}
\newcommand{\Hab}{\Ha/\Hb}
\newcommand{\FIRS}{ALMA\,--\,NIRSpec}
\newcommand{\ZOH}{12+log$_{10}$(O/H)}

\begin{document}

   \title{The ionised interstellar medium of DSFGs revealed by JWST/NIRSpec and ALMA: Super-solar metallicity, low ionisation parameters and, typical electron densities}

    \author{Steven Gillman,\inst{1,2}\orcidlink{0000-0001-9885-4589} 
          Kei Ito,\inst{1,2}\orcidlink{0000-0002-9453-0381} 
          Francesco Valentino, \inst{1,2}\orcidlink{0000-0001-6477-4011} 
          Gabe Brammer, \inst{1,3}\orcidlink{0000-0003-2680-005X} 
          Pablo Araya Araya, \inst{1,2}\orcidlink{0000-0003-2860-5717} 
          Georgios Magdis,\inst{1,2}\orcidlink{0000-0002-4872-2294} 
          Ugnė Dudzevičiūtė,\inst{1,2}\orcidlink{0000-0003-4748-0681}
          Aswin P. Vijayan,\inst{4}\orcidlink{0000-0002-1905-4194}
          Minju Lee,\inst{1,2}\orcidlink{0000-0002-2419-3068} 
          Bitten Gullberg,\inst{1,2}\orcidlink{0000-0002-4671-3036} 
          Daniel Ceverino, \inst{5,6}\orcidlink{0000-0002-8680-248X}         
          Andreas L. Faisst, \inst{7}\orcidlink{0000-0002-9382-9832} 
          Seiji Fujimoto, \inst{8,9}\orcidlink{0000-0001-7201-5066}
          Thomas R. Greve, \inst{1,2}\orcidlink{0000-0002-2554-1837} 
          Rashmi Gottumukkala, \inst{1,3} \orcidlink{0000-0003-0205-9826}
          Chandana Chandrashekhar, \inst{1,2} \orcidlink{0009-0004-8166-1106}
          Michaela Hirschmann, \inst{10} \orcidlink{0000-0002-3301-3321}
          Shuowen Jin, \inst{1,2}\orcidlink{0000-0002-8412-7951} 
          Christian Kragh Jespersen, \inst{11}\orcidlink{0000-0002-8896-6496}  
          Takumi Kakimoto, \inst{12,13}\orcidlink{0000-0003-2918-9890}
          Mariko Kubo, \inst{14,15} \orcidlink{0000-0002-7598-5292}
          Peter Laursen, \inst{1,3} \orcidlink{0000-0003-4207-0245}
          Masato Onodera, \inst{12,16} \orcidlink{0000-0003-3228-7264}
          Antonio Pensabene, \inst{1,2}\orcidlink{0000-0001-9815-4953}  
          Francesca Rizzo, \inst{17}\orcidlink{0000-0001-9705-2461}
          John R. Weaver, \inst{18}\orcidlink{0000-0003-1614-196X}
          Po-Feng Wu, \inst{19,20}\orcidlink{0000-0002-9665-0440}  
          }

       \institute{
        Cosmic Dawn Center (DAWN), Denmark\\
        \email{srigi@space.dtu.dk}
        \and
        DTU-Space, Elektrovej, Building 328 , 2800, Kgs. Lyngby, Denmark
        \and 
         Niels Bohr Institute, University of Copenhagen, Jagtvej 128, DK-2200, Copenhagen N, Denmark
        \and 
        Astronomy Centre, University of Sussex, Falmer, Brighton BN1 9QH, UK
        \and 
        Departamento de Fisica Teorica, Modulo 8, Facultad de Ciencias, Universidad Autonoma de Madrid, 28049 Madrid, Spain
        \and
        CIAFF, Facultad de Ciencias, Universidad Autonoma de Madrid, 28049 Madrid, Spain
        \and 
        Affiliation: IPAC, California Institute of Technology, 1200 E. California Blvd. Pasadena, CA 91125, USA
        \and 
        David A. Dunlap Department of Astronomy and Astrophysics, University of Toronto, Toronto, ON M5S 3H4, Canada
        \and
        Dunlap Institute for Astronomy and Astrophysics, University of Toronto, Toronto, ON M5S 3H4, Canada
        \and 
        Institute of Physics, Laboratory for galaxy evolution, Observatory of Sauverny, Chemin Pegasi 51, 1290 Versoix, Switzerland
        \and
        Department of Astrophysical Sciences, Princeton University, Princeton, NJ 08544, USA
        \and 
        Department of Astronomical Science, The Graduate University for Advanced Studies, SOKENDAI, 2-21-1 Osawa, Mitaka, Tokyo 181-8588, Japan
        \and 
        National Astronomical Observatory of Japan, 2-21-1 Osawa, Mitaka, Tokyo 181-8588, Japan
        \and 
        Department of Physics and Astronomy, School of Science, Kwansei Gakuin University, 1 Gakuen Uegahara, Sanda, Hyogo 669-1330, Japan
        \and
        Astronomical Institute, Tohoku University, 6-3, Aramaki, Aoba, Sendai, Miyagi 980-8578, Japan
        \and 
        Subaru Telescope, National Astronomical Observatory of Japan, National Institutes of Natural Sciences (NINS), 650 North A’ohoku Place, Hilo, HI 96720, USA
        \and  
        Kapteyn Astronomical Institute, University of Groningen, Landleven 12, 9747 AD, Groningen, The Netherlands
        \and 
        MIT Kavli Institute for Astrophysics and Space Research, 70 Vassar Street, Cambridge, MA 02139, USA
        Graduate Institute of Astrophysics, National Taiwan University, Taipei 10617, Taiwan
        \and 
        Department of Physics and Center for Theoretical Physics, National Taiwan University, Taipei 10617, Taiwan
        }
        
   \date{\today}

  \abstract{We present a detailed study of the observed-frame near-infrared (2\,--\,4\,\um) JWST/NIRSpec spectra of 48 high-redshift ($z$\,=\,2.53$^{+1.32}_{-0.70}$) galaxies detected with ALMA at $>$3$\sigma$. Through a multi-wavelength SED analysis we establish the sample has a median stellar mass of $\log_{10}$(M$_{\rm \ast}$[M$_{\odot}$])\,=\,10.8\,$\pm$\,0.1 and dust mass of $\log_{10}$(M$_{\rm d}$[M$_{\odot}$])\,=\,8.7\,$\pm$\,0.1,  covering a broad range of far-infrared luminosity ($\log_{10}$(L$_{\rm FIR}$[L$_{\odot}]$)\,=\,10.9\,--\,12.7). 
  Analysing the rest-frame optical emission-line properties, we discern the majority of the sources show no indication of AGN activity with 40\% having either X-ray counterparts (L$_{\rm Xc}$\,$>$\,10$^{42}$\,erg/s), elevated optical (\OTHb, \NH)  emission-line ratios or broad (FWHM\,$>$\,800\,km/s) \Ha\ emission line profiles, although we note this is a lower limit given the stochastic placement of NIRSpec slits. We establish the sample has a median gas-phase metallicity of \ZOH\,=\,8.71\,$\pm$\,0.02, as derived from the \NH\ ratio, with 
 the more far-infrared luminous ($\log_{10}$(L$_{\rm FIR}$[L$_{\odot}]$)\,$>$12.0)  galaxies exhibiting 0.15\,$\pm$\,0.03\,dex higher metallicity at their epoch than predicted by the fundamental metallicity relation.
 From the \sii\ emission-line doublet ratio, we measure a median electron density for the sample of $\log_{10}(n_{\rm e}$[cm$^{-3}$])\,=\,2.53\,$\pm$\,0.07, consistent with less-massive, star-forming, galaxies at the same epoch. Finally, for a subsample of nine sources with \oii\ and \Hb\ detections (median $\log_{10}$(L$_{\rm FIR}$[L$_{\odot}]$)\,=\,11.81\,$\pm$\,0.15), we derive a median observed (dust-uncorrected) ionisation parameter of $\log_{10}(U)$\,=\,$-$2.84\,$\pm$\,0.06, suggesting our sample fall below the ionisation parameter \Ha\ star-formation rate surface density ($\Sigma_{\rm H\alpha,SFR}$) relation of typical star-forming galaxies.
Our results indicate that luminous far-infrared galaxies are massive, chemically evolved, systems that appear to deviate from the standard dust and metal production equilibrium observed in less obscured galaxies. This study demonstrates the powerful synergy of JWST and ALMA in unveiling the nature of dusty star-forming galaxies, and highlights the need for a NIRSpec survey of uniformly selected, massive, dust obscured, galaxies to fully characterise their interstellar medium.} 

\titlerunning{Ionised ISM Conditions in DSFGs from JWST and ALMA}
\authorrunning{S. Gillman et al.}

\date{Received 24 February 2026 / Accepted 25 September 2026}

\maketitle
\section{Introduction}

In the last few decades, large photometric surveys have revealed that up to 70\% of the co-moving star formation rate density at 0.5$<$\,$z$\,$<$\,4 is obscured by dust \citep[e.g.][]{Elbaz2001,Schreiber2015}, with this fraction declining at higher redshifts \citep[e.g.][]{Algera2023,Martis2025}. At Cosmic Noon ($z$\,$\simeq$\,1\,--\,4) the most extreme examples of galaxies with heavily obscured emission are sub-millimetre bright galaxies (SMGs) \citep[e.g.][]{Blain2002, Casey2014,Dud2020,Simpson2020,Liao2024} but there are also significant contributions from fainter ($<$1\,mJy) sub-millimetre galaxies \citep[e.g.][]{Cowie2002, Clements2008, Fujimoto2016, Aravena2016,Tomoko2021,Smail2026}, that together are collectively known as dusty star-forming galaxies (DSFGs). For a full review see \citet{Casey2014}.

The high star formation rates (SFR\,$\sim$\,10$^{2-3}$\,M$_{\odot}$\,yr$^{-1}$) and large stellar masses ($>$10$^{10}$M$_{\odot}$) of DSFGs has proved challenging to reproduce in theoretical galaxy formation models \citep[e.g.,][]{Baugh2005,Swinbank2008,Hayward2013,Mcalpine2019}, although more recent attempts have been more successful \citep[e.g.,][]{Lovell2022,Lower2022,Cochrane2023}. The models suggest that the high star formation rates
in this population are driven by a mix of secular instabilities in gas-rich discs and dynamical triggers due to minor and
major mergers  \citep[e.g.,][]{Mcalpine2019}.

However, in contrast high-resolution observational studies of the dust continuum in DSFGs \citep[e.g.,][]{Simpson2015,Ikarashi2015,Fujimoto2017} has revealed compact disk-like structures (i.e., $R_{\rm e}$\,$\sim$\,1\,--\,2\,kpc, S\'ersic $n$\,$\approx$\,1) as well as evidence for a fainter more extended component with $R_{\rm e}$\,$\simeq$\,4\,kpc \citep[e.g.,][]{Gullberg2019,Ivison2020}. In addition, resolved kinematic studies using molecular and atomic fine structure emission lines in the rest-frame far-infrared and sub-/millimetre have uncovered disk-like kinematics for at least a significant fraction of the population \citep[e.g.,][]{Hodge2012,Chen2017,Lelli2021,Rizzo2021,Amvrosiadis2023}. More recent high resolution studies with JWST have further revealed remarkably uniform, disc-dominated, morphologies at rest-frame near-infrared (1\um) wavelengths \citep[e.g.][]{Gillman2024,Mckinney2024,Bodansky2025}, with only the most extreme, high redshift ($z$\,$>$\,3) DSFGs, showing high fractions of mergers and interactions \citep[e.g.][]{Hodge2025}. This suggests for the bulk of the DSFG population, their evolution is secularly driven and  not dominated by major mergers as previously suggested.
   
This raises the question then what is driving the extreme, dust-obscured, often star-bursting, nature in high-redshift DSFGs, given their apparent secular (i.e non-major merger) dominated evolution. Less dust-obscured star-forming galaxies at the same epoch, follow tight scaling relations, such as the main sequence of star formation \citep[e.g.][]{Speagle2014,Popesso2023}, the mass metallicity relation \citep[e.g.][]{Tremonti2004, Sanders2021} and fundamental metallicity relation \citep[e.g.][]{Mannucci2010,Curti2020}, that represent the equilibrium status between ongoing gas consumption and star formation, prior star-formation (stellar mass) alongside metal enrichment and dust production processes (see \citealt{NFS2020} for a full review). It is often assumed that dust-obscured galaxies exhibit the same equilibrium between star-formation and dust and metal enrichment, with many studies employing the scaling relations to indirectly infer gas-phase metallicities \citep[e.g.][]{Liu2019,Donevski2020,Kokorev2021,Gomez2022}. However, given the extreme nature of DSFGs, with high stellar masses and star formation rates, whether such an assumption is valid is not well defined.

Whilst constraining the stellar mass and star-formation rates, and thus relation to the main sequence of star formation, of dust obscured galaxies has been possible for many years \citep[e.g.][]{Swinbank2014,Dunlop2017,Long2023}, with the most heavily dust obscured systems elevated specific star formation rates placing them in the starburst regime \citep[e.g.][]{Silverman2015,Dud2020,Puglisi2021,Faisst2025}, measuring the galaxies metallicity has remained a challenge. To constrain the metallicity of the interstellar medium 
we would ideally use the far-infrared fine structure lines (e.g.\,[O \textsc{i}] 63\um, [Si \textsc{ii}] 34\um, [N \textsc{iii}] 57\um), which are unaffected by dust obscuration. However, at $z$\,$<$\,4, ALMA can not detect the multiple line diagnostics required \citep[e.g.][]{Decarli2025}. Prior studies using the Herschel space telescope achieved detections in stacked low signal to noise spectra of the far-infrared fine structure lines \citep[e.g.][]{Coppin2012,Wardlow2017,Wilson2017,Rigopoulou2018,Bakx2026} or utilised lensing \citep[e.g.][]{Zhang2018}, to reveal on average super solar metallicities. 

Another option is to use rest-frame optical lines, in combination with strong-line calibrations \citep[e.g.][]{Maiolino2019}, to derive gas-phase metallicities and other interstellar medium properties. Although, these should not be interpreted as direct equivalents of those inferred from far-infrared fine-structure lines, as the two sets of diagnostics probe gas at different characteristic densities and physical conditions \citep[e.g.][]{Harikane2025,Usui2025}. Detecting rest-frame optical emission lines in high-redshift ($z$\,$>$\,1) in dust-obscured galaxies is traditionally very challenging due to the their inherent optical faintness. Limitations from ground-based instrumentation sensitivity often results in low signal to noise spectra dominated by atmospheric affects (i.e. OH skylines) with limited wavelength coverage in a subsample of the brightest, least dust obscured, sources \citep[e.g.][]{Swinbank2004,Ala2012,Del2013,Casey2017,Tomoko2021, Shapley2020,Hogan2021,Birkin2023}.

With the advent of JWST/NIRSpec it is now possible to detect the rest-frame optical emission lines at high-redshift in dust-obscured galaxies at high signal to noise and resolution. Early JWST studies utilised the NIRSpec/PRISM mode to reveal the diverse rest-frame optical spectroscopic properties of a small sample of far-infrared bright galaxies \citep[e.g.][]{Cooper2025,Price2025}, but lacked the spectroscopic resolution to resolve key line ratios, and thus infer interstellar medium properties. A more recent study by \citet{Kiyota2026} utilised medium resolution (R\,$\sim$\,1000) NIRSpec/MSA spectroscopy from the JWST Advanced Deep Extragalactic Survey \citep[JADES;][]{Eisenstein2023} program of 16 sub-millimetre faint ($S_{\rm B6}$\,=\,0.1\,--\,0.99\,mJy) ALMA sources from the ALMA Spectroscopic Survey in the Hubble Ultra Deep Field \citep[ASPECS;][]{Walter2016} survey. They identify nominal metallicities and electron densities suggesting the galaxies follow the same scaling relations as typical star-forming galaxies at the same epoch. However, asides from a relatively small number of sources, the sample does not include the most dust-obscured, bright ($>$1\,mJy), sub-millimetre sources, and therefore the nature of these extreme sources remains unclear.

To this end, in this paper, we present an analysis of the rest-frame optical spectroscopic properties of every known ($>$3$\sigma$) ALMA detected sources currently observed with NIRSpec/MSA. In Section \ref{Sec:Sample} we define the initial sample of \FIRS\ galaxies that we use in our analysis whilst in Section \ref{Sec:Obs} we present the observations, data reduction and analysis undertaken on these sources, before constructing a robust final sample. In Section \ref{Sec:Results} we give our results and discuss their implications before summarising our main conclusions in Section \ref{Sec:Conc}. Throughout the paper, we assume a $\Lambda$CDM cosmology with $\Omega_{\rm m}$\,=\,0.3, $\Omega_{\Lambda}$\,=\,0.7, and $H_0$\,=\,70\,$\mathrm{km\,s^{-1}\,Mpc^{-1}}$. All quoted magnitudes are on the AB system \citep{Oke1983}, and stellar masses are calculated assuming a Kroupa initial mass function (IMF) \citep{Kroupa2002}. All uncertainties on median values are derived from bootstrapped resampling.

\begin{figure*}[!htbp]
    \centering
    \includegraphics[width=\linewidth]{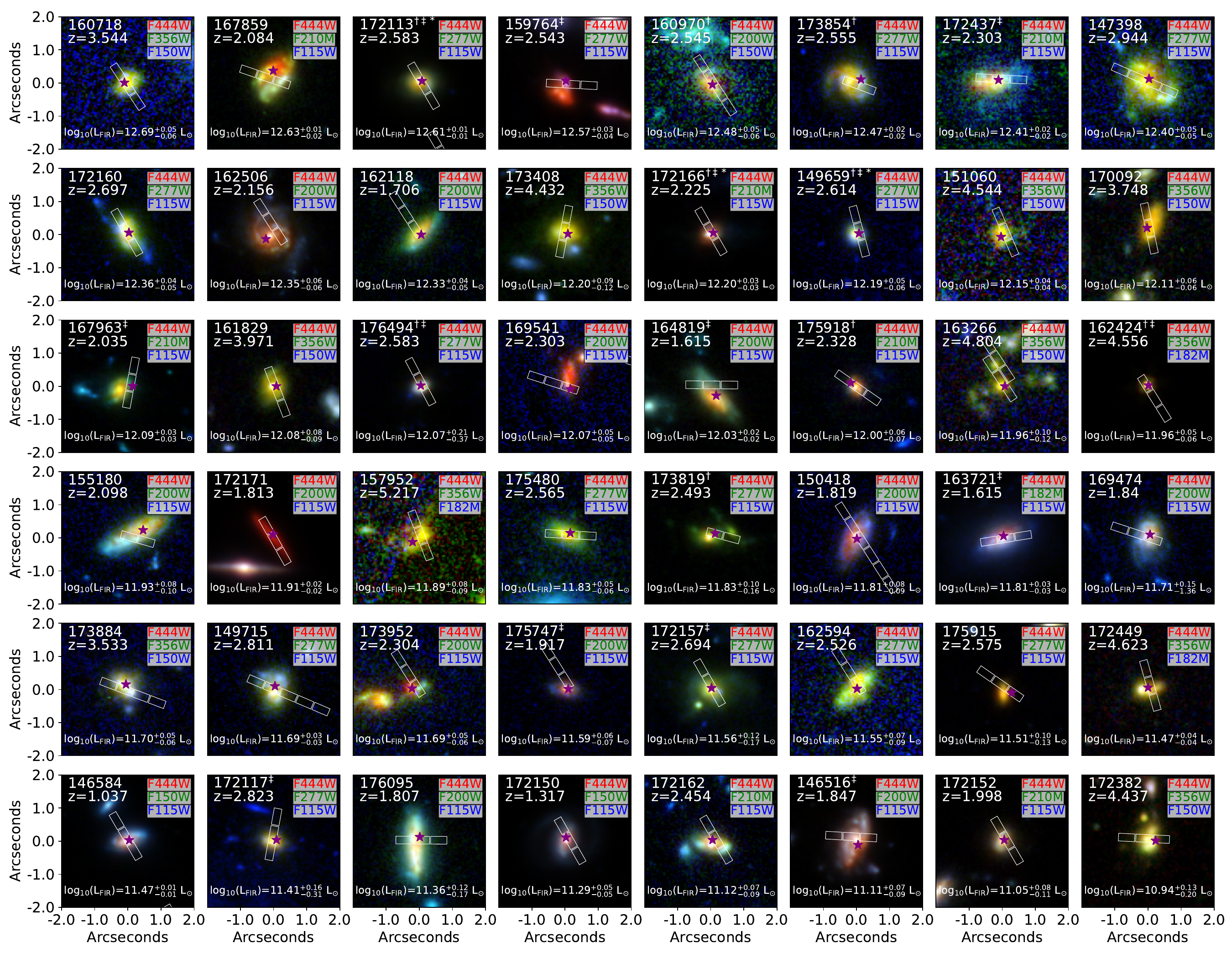}
    \caption{JWST/NIRCam false colour images for the final sample of 48 \FIRS\ galaxies ranked in descending far-infrared luminosity (L$_{\rm FIR}$). For each source we show the 4\farcs{0}$\,\times\,$4\farcs{0} cutout as labelled with the DJA-ID, spectroscopic redshift, far-infrared luminosity, ALMA peak flux centroid (purple star) and NIRSpec MSA slits (white boxes) for one dither position. For each source we select the NIRCam filter that corresponds to rest-frame 2\um\ (R, near-infrared), 0.7\um\,(G, optical) and   0.3\um\ (B, ultra-violet). Sources identified as AGN (see Sec. \ref{Sec:AGN}) are highlighted, where $\dagger$ indicates broad emission lines, $\ddagger$ indicates an X-ray counterpart and $*$ highlights elevated optical-line ratios.}
    \label{fig:RGB}
\end{figure*}

\section{Sample selection}\label{Sec:Sample}

To constrain the ionised interstellar medium properties (e.g. metallicity, electron density, ionisation parameter) of known ALMA sources, we require at least median resolution (R\,$\gtrsim$\,1000) spectroscopy that samples the rest-frame optical to near-infrared wavelength regime. This provides both the spectral coverage for key emission line features (e.g. \Ha, \Hb, \oiii) as well as the resolution ($\leq$\,127\,km\,s$^{-1}$) to resolve line complexes such as \NH. 

Thus, to build a sample of galaxies with both median resolution spectroscopy  and prior far-infrared detections, we use observations from the DeepDive Survey (PID \#3567, PI: Valentino). For full details of the DeepDive Survey see \citet{Ito2025}. In brief, DeepDive is a JWST Cycle-2 program using the NIRSpec Mirco-Shutter Array \citep[MSA;][]{Jakobsen2022}, targetting quiescent galaxies at $z$\,$\simeq$\,3\,--\,4  with the median resolution G235M/F170LP filter grating pair. In addition to the primary targets, additional filler targets of known far-infrared detected sources were also included in the MSA masks, in total giving spectra for 555 galaxies across a broad range of redshift (see \citet{Ito2025} for details).

We utilise this sample from DeepDive, in addition to the full DAWN JWST Archive \citep[DJA\footnote{\url{https://doi.org/10.5281/zenodo.15472354}};][]{Valentino2023,Heintz2025} v4.4 that contains all public NIRSpec spectra to date. Next, 
we compile a list of known ALMA detected sources. 
Specifically, we utilise the ECOlogy for Galaxies using ALMA archive and Legacy surveys \citep[ECOGAL;][]{Lee2025} database in the UKIDSS Ultra-DEEP Survey \citep[UDS;][]{Lawrence2007} extragalactic field, in addition to 
A3COSMOS\footnote{v20250312\,\url{https://uni-bonn.sciebo.de/s/0LrfuA14s2YP3cz?path=\%2F}} \citep{Liu2019} and A3GOODS-S\footnotemark[2] \citep{A3GOODS2024} in the Cosmic Evolution Survey \citep[COSMOS;][]{Scoville2007} and Great Observatories Origins Deep Survey – South \citep[GOODS-S;][]{Giavalisco2004} extragalactic fields respectively. The ECOGAL database provides a list of ALMA detected sources
in regions that overlap with the HST Cosmic Assembly Near-infrared Deep Extragalactic Legacy Survey \citep[CANDELS;][]{Grogin2011} footprint (see \citealt{Lee2025} for details), whilst A3COSMOS and A3GOODS-S present all sources detected in the ALMA archive in their respective fields. We further make use of a list of previously characterised bright ($S_{870\mu m}$\,$>$1\,mJy) sub-millimetre galaxies in the literature with high-resolution ($<$\,0\farcs{6}) counterparts, as presented in \citet{Gillman2023}, which includes 
ALMA-detected SCUBA-2 sources from \citealt{Dud2020}. 

Using this combined list, we cross match the position of the NIRSpec slit to ALMA source positions with a search radius of $r$\,=\,0\farcs{5}. In doing so, we both capture any NIRSpec source with a far-infrared ALMA detection, as well as known sub-millimetre bright galaxies that fall outside the HST-CANDELS footprint. In total we identify 153 unique galaxies with NIRSpec spectra, of which 32 have PRISM spectra and 121 have median resolution (or higher) grating spectra.

\section{Data reduction and analysis}\label{Sec:Obs}

Having defined an initial sample of 121 ALMA sources with median, or higher, resolution NIRSpec observations, we now first analyse their photometric properties, at both near and far-infrared wavelengths, in addition to their spectroscopic properties, to establish the presence of key rest-frame optical emission lines required to reveal their interstellar medium properties, before defining a robust final sample.

\subsection{UV-NIR photometry and SED fitting}

We use the public database of existing imaging in well studied extragalactic fields from DJA (for details see \citealt{Valentino2023}) that incorporates deep multi-band HST (ACS + WFC3) observations from the HST-CANDELS program and JWST (NIRCam + MIRI) observational programs such as; The Public Release IMaging for Extragalactic Research \citep[PRIMER;][]{Dunlop2021}, JADES Survey \citep[][]{Eisenstein2023}. For full details of the observational imaging programs we refer the reader to DJA\footnote{\url{https://dawn-cph.github.io/dja/imaging/v7/}}. Nearly 78\% (94/121) of the sources fall within the either UDS field (50 sources) or GOODS-S (44 sources) whilst the remaining 27 sources come from the COSMOS field extragalactic field. Figure \ref{fig:RGB} shows the NIRCam false colour images for the final sample as defined in Section \ref{Sec:Final}.

For each source we adopt the photometry from DJA. As detailed in \citep{Valentino2023}, we measure the photometry using the pythonic version of the Source Extractor code \texttt{sep v1.2.1} \citep{Bertin1996, Barbary2016}. We use a combined NIRCam long-wavelength (LW) image as a detection image and measure the flux in circular apertures with a diameter of 0\farcs{5} and corrected to "total" values with Kron apertures \citep{Kron1980} which were computed on the LW combined detection image.
To derive the stellar mass of the sources, we model their SEDs using the Bayesian Analysis of Galaxies for Physical Inference and Parameter EStimation  \citep[\texttt{bagpipes};][]{Carnall2018} code, with a double power-law star formation history and a \citet{Calzetti2000} dust attenuation law. We adopt the \citet{Bruzual2003} stellar population models with a \citet{Kroupa2002} initial mass function. We adopt uniform  priors for the stellar mass and \av\ with a logarithmic prior on the metallicity (see \citealt{Ito2025} for full details)\footnote{\tcr{We note for sources with non-negligible AGN contributions, the stellar masses may represent upper limits in our analysis.}}. Finally, we fixed the redshift to the spectroscopic redshift as estimated from \texttt{msaexp} (see Sec. \ref{Sec:Spec}). The SED properties of the final sample (Section \ref{Sec:Final}) is presented in Appendix \ref{App: SED_table}.

\subsection{MIR-radio photometry and SED fitting}\label{Sec:BB}

In addition to the ALMA observations from which our sample was defined (see Table \ref{App:Selection}), we compile the mid-to far-infrared photometry for our sources from the literature. For sources in the UDS field we adopt the Herschel de-blended photometry catalogue as utilised in \citep{Swinbank2014,Dud2020}, as well as the ALMA photometry from ECOGAL \citep{Lee2025}. For COSMOS, we use an updated super-deblended catalogue from \citet{Jin2018}, whilst for GOODS-S we adopt the infrared photometry from \citet{Magnelli2014}. We identify the photometry for each source by cross matching the photometry catalogues with the NIRSpec source position (see Table \ref{App:Selection}), finding all matches within 1\farcs{1}. We enforce a detection criteria of 3$\sigma$ in a given band, defining 3$\sigma$ upper limits for all fluxes detected at a lower significance.
We remove any photometry that samples rest-frame blueward of 40\um\, as this requires multi-temperature blackbody models that can incorporate the warm dust emission of the sources \citep[e.g.][]{Draine2007,Cunha2008}, which is beyond the scope of this paper.

\begin{figure*}[!htbp]
    \centering
    \includegraphics[width=0.345\linewidth,trim={0cm 0cm 0cm 0cm},clip]{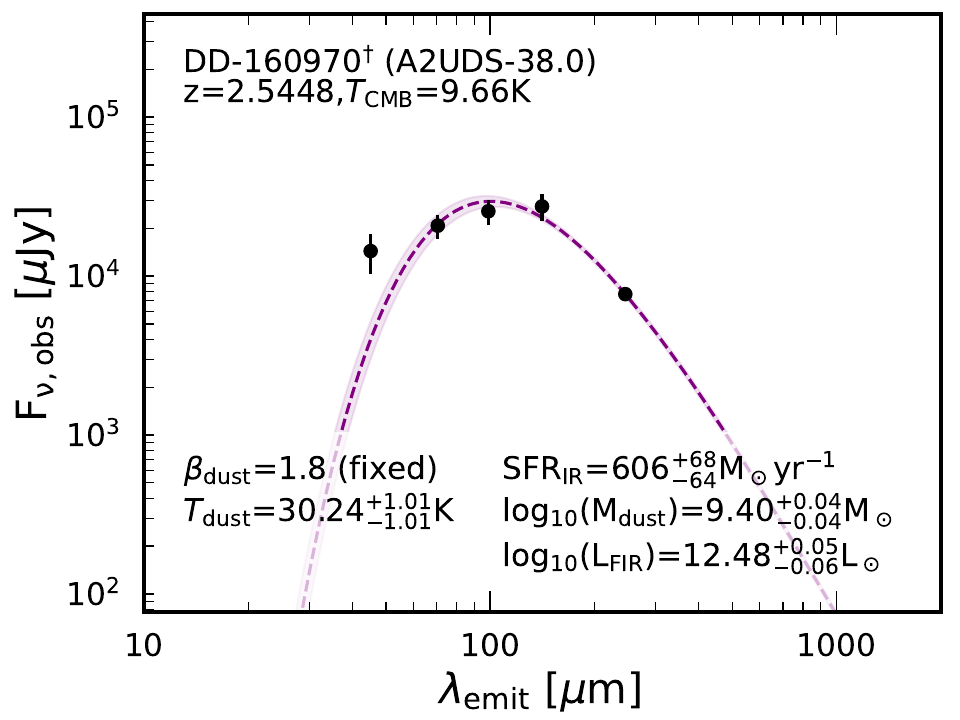}
    \includegraphics[width=0.322\linewidth,trim={1cm 0cm 0cm 0cm},clip]{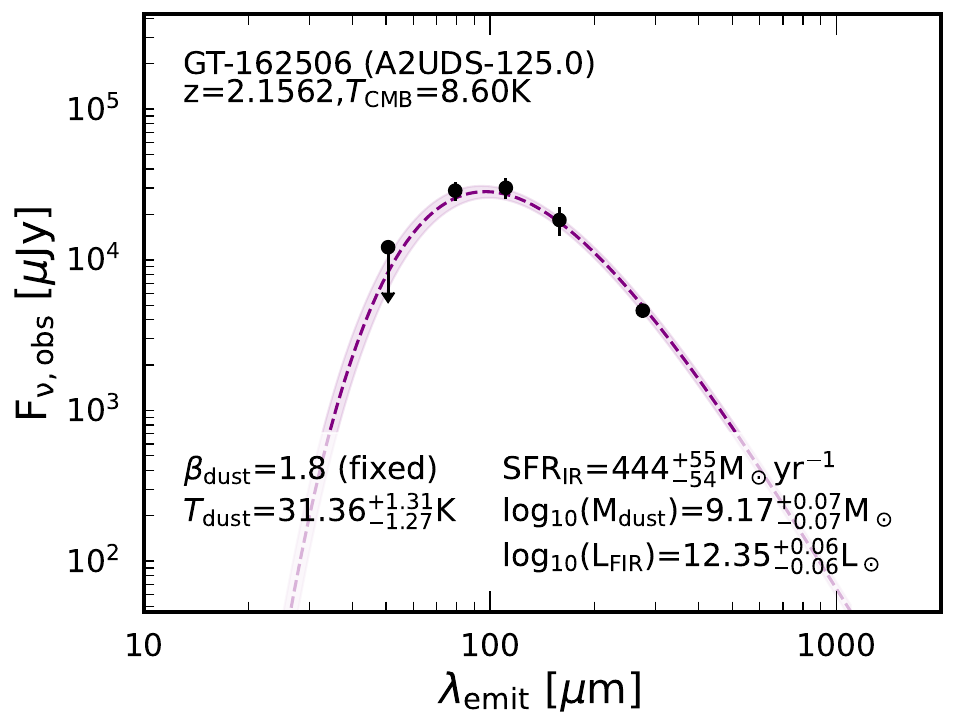}
    \includegraphics[width=0.322\linewidth,trim={1cm 0cm 0cm 0cm},clip]{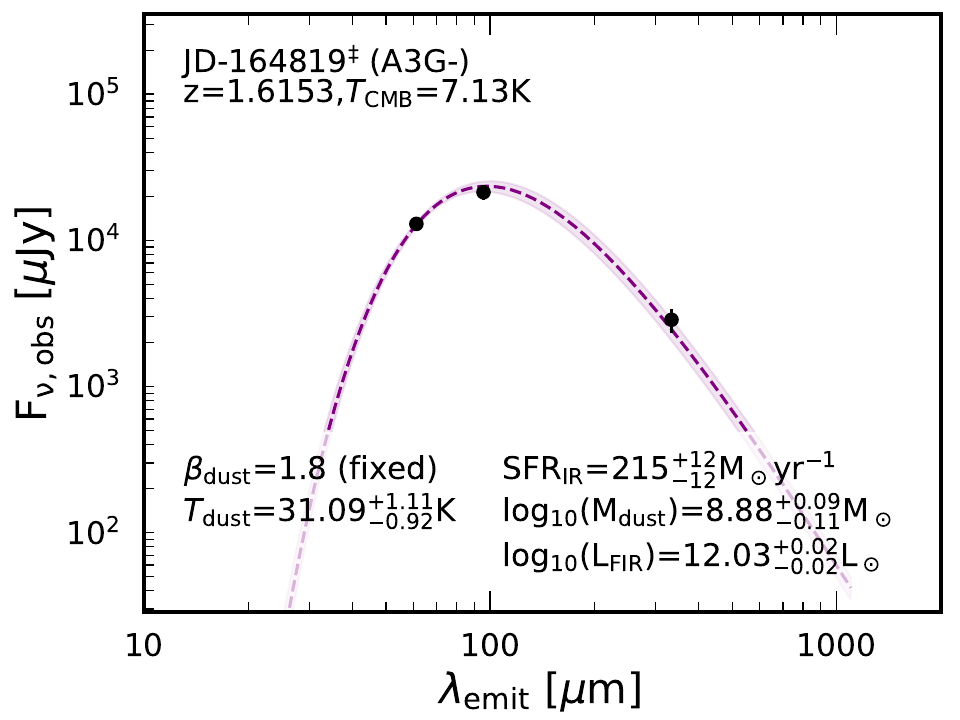}
     \caption{Three examples of the \texttt{mercurius} far-infrared modified blackbody modelling. For each source show the photometry (black points; upper limits shown as arrows) and best-fit blackbody curve (purple-dashed line) and 1$\sigma$ uncertainty (purple shaded region). We report the best-fit parameter in each panel, as well as NIRSpec (and ALMA) source ID, redshift and CMB temperature. Sources identified as AGN (see Sec. \ref{Sec:AGN}) are highlighted, where $\dagger$ indicates broad emission lines, $\ddagger$ indicates an X-ray counterpart and $*$ highlights elevated optical-line ratios. The SEDs, and MCMC corner plots, of the full sample are given in Section \ref{App:FIR}.}
    \label{fig:FIR_SEDS}
\end{figure*}

To derive the far-infrared SED properties of the sources we utilise the Multimodal Estimation Routine for the Cosmological Unravelling of Rest-frame Infrared Uniformised Spectra \citep[\texttt{mercurius};][]{Witstok2022} code, which performs a Bayesian least-squares fitting of a modified blackbody fitting to far-infrared SED, examples of which are shown in Figure \ref{fig:FIR_SEDS}. Given the well known degeneracy between dust temperature ($T_{\rm d}$) and far-infrared slope ($\beta$), we fix the far-infrared slope to $\beta$\,=\,1.8, consistent with values derived for DSFGs from Herschel and ALMA observations \citep[e.g.][]{Magnelli2014,daCunha2015,Ward202_dust,Bing2024}. For each source we fixed the redshift to the spectroscopic redshift as estimated from \texttt{msaexp} (see Sec. \ref{Sec:Spec}). We further assume  optically thin dust and adopt a flat prior on the dust mass in the range $\log_{10}$(M$_{\rm d}$[M$_{\odot}$])\,=\,4.0\,--\,12.0. For the dust temperature, we used the default gamma distribution, shifted to the temperature of the cosmic microwave background, with shape parameter fixed to 1.5, with a maximum dust temperature of 150\,K. For 25 sources with a single far-infrared detection (excluding limits), we further fix the dust temperature to $T_{\rm d}$\,=\,30\,K. We adopt the built-in dust emissivity coefficient $\kappa$\,=\,$\kappa_0(\nu/\nu_0)^{\beta}$ with $\kappa_0$\,=\,8.94\,cm$^{2}$\,g$^{-1}$ at $\nu_0$\,=\,1900\,GHz ($\sim$158\um\ \citealt{Hirashita2014}). 
Details of the \texttt{mercurius} blackbody fits, including MCMC corner plots, for the final sample (Sec. \ref{Sec:Final}), are given in Section \ref{App:FIR}, whilst the derived physical parameters are given in Appendix \ref{App: SED_table}.


\subsection{Rest-frame optical morphology} \label{Sec:Sizes}

To constrain the rest-frame optical morphology of the ALMA detected galaxies we use \texttt{PySersic} \citep{PySersic}. \texttt{PySersic} is a python-based Bayesian inference code which fits parametric (S\'ersic) models to galaxy light distributions. For each galaxy, we fit the S\'ersic profiles to the available JWST NIRCam wide-band imaging in 4\farcs{8}\,$\times$\,4\farcs{8} size cutouts, using point spread functions PSFs derived from \texttt{PSFEx} \citep{Bertin2011} using the same procedure as \citet{Genin2025}.
Nearby sources are detected down to a 1.5$\sigma$ level and masked using the segmentation map in the fitting process. We then derive the rest-frame optical (0.5\um) size by selecting the NIRCam long-wavelength filter with rest-frame wavelength closest to 0.5\um\, in a similar methodology to \citet{Gillman2023}. The derived sizes and S\'ersic indices, for the final sample (see Sec. \ref{Sec:Final}), are given in Appendix \ref{App: SED_table}.

\subsection{Spectroscopy}\label{Sec:Spec}

To constrain the ionised interstellar medium properties of the far-infrared detected NIRSpec sources, we first measure the continuum level in the spectra before flux calibrating and quantifying the emission-line properties.

\subsubsection{Continuum and flux calibration}

To model the continuum we run \texttt{msaexp} \citep{Brammer2023, Graaf2025}. Using a non-parametric approach, with a combination of splines and Gaussian models for the emission lines, \texttt{msaexp} produces a robust estimate of the continuum level, in addition to the noise spectrum, as a function of wavelength. For our analysis we use 13 knot-splines to model the continuum. The \texttt{msaexp} pipeline further extends the standard wavelength coverage of the NIRSpec spectra, to the full red extension using a combination of long-pass filter and the projection on the detector (see \citealt{Valentino2025} for details). 

\begin{figure*}[!htbp]
    \centering
    \includegraphics[width=\linewidth]{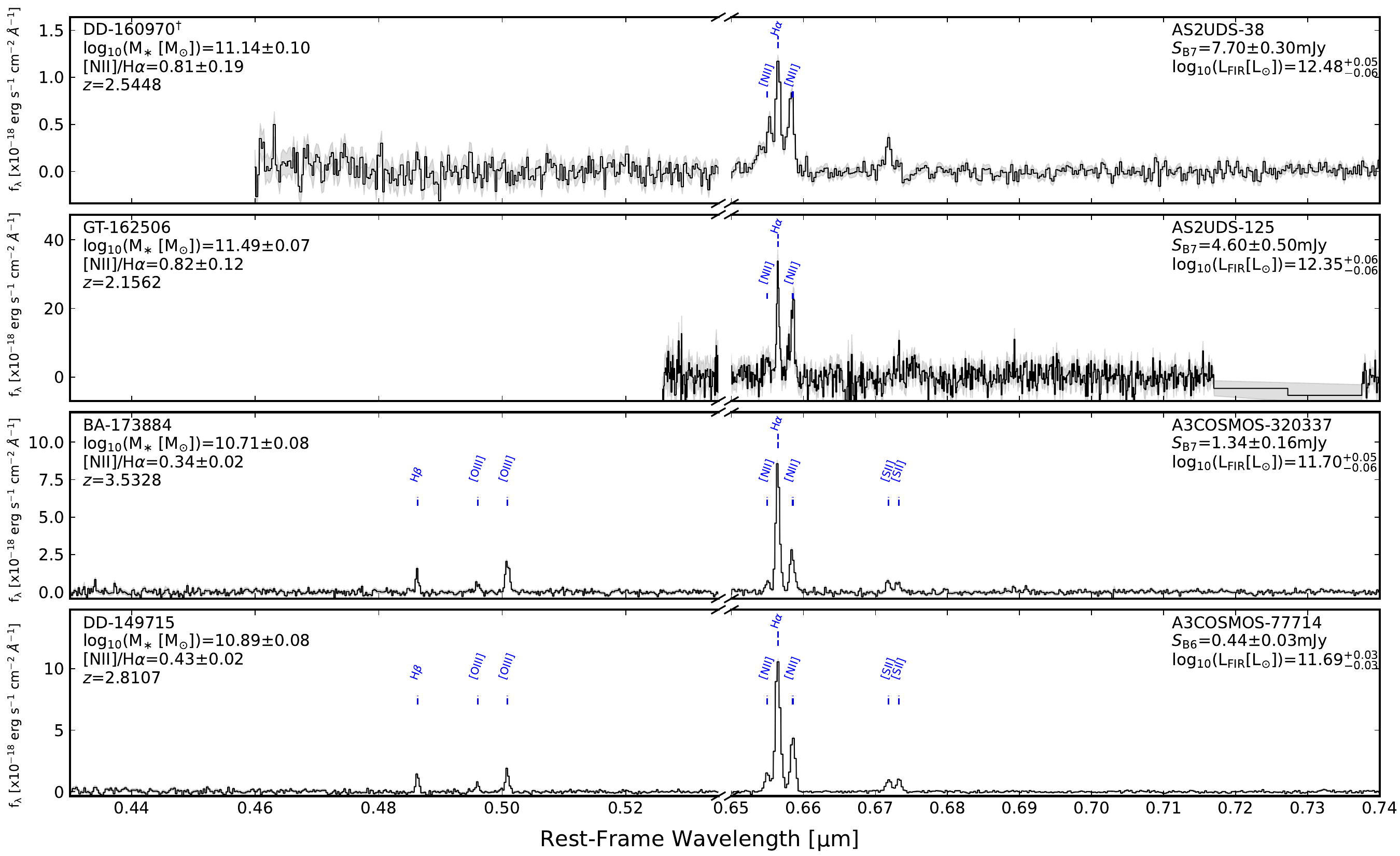}
    \caption{JWST/NIRSpec spectra for four galaxies in the sample across a range of far-infrared luminosity. For each source we show the \tcr{continuum subtracted} NIRSpec spectra (and 1$\sigma$ uncertainty) from rest-frame 0.36\,--\,0.54\um\ highlighting the \OTHb\ complex and  0.65\,--\,0.74\um\ highlighting the \NH\ complex, with detected ($>$3$\sigma$) emission lines labelled. We further indicate the NIRSpec-ID, \texttt{bagpipes} derived stellar mass, observed \NH\ ratio, spectroscopic redshift as well as ALMA-ID, ALMA flux density and far-infrared luminosity of each source. Sources identified as AGN (see Sec. \ref{Sec:AGN}) are highlighted, where $\dagger$ indicates broad emission lines, $\ddagger$ indicates an X-ray counterpart and $*$ highlights elevated optical-line ratios.}
    \label{fig:Spectra}
\end{figure*}

As described in \citet{Ito2025}, we utilise the available multi-wavelength photometry to flux calibrate the spectra. We  model any potential flux losses using a polynomial function, whose order reflects the number of available photometric bands.\footnote{We note that given the stochastic placement of MSA slits, driven by the archival nature of our sample, the spectral properties may not always represent the full galaxy integrated properties.} 
We further apply a wavelength-dependent correction that accounts for parts of the spectrum beyond the nominal cut-off (see \citealt{Ito2025} for full details.). Furthermore, \texttt{msaexp} defines the spectroscopic redshift of the source through the detection of multiple emission lines. We show examples of the rest-frame, continuum subtracted, spectra in Figure \ref{fig:Spectra}, with the spectra for the final sample (Sec. \ref{Sec:Final}) given in Section \ref{App:FIR}.

\subsubsection{Emission-line fitting}\label{Sec:Fitting}

Next, we perform the emission-line fitting to the continuum subtracted spectrum using the \texttt{mpfit}\footnote{\url{https://millenia.cars.aps.anl.gov/software/python/mpfit.html}} python code that employees the Levenberg-Marquardt technique to solve the least-squares problem. We use \texttt{mpfit} instead of \texttt{msaexp} as it allows more flexibility in the modelling of multiple Gaussian profiles with interconnected properties. For instance to model line complexes such as  \Ha-\nii-\sii\ or \oiii-\Hb, we use multiple Gaussian profiles with consistent redshifts and emission-line widths, whilst the amplitude of each emission line is free to vary.  
For single emission-line features we fit a single Gaussian model with emission-line width, amplitude and redshift free to vary. For all emission-lines the minimum width is fixed by the instrumental broadening of the NIRSpec MSA instrument as a function of wavelength. We adopt the post-flight improved resolution of NIRSpec, as estimated by recent studies to be of order $\sim$30\%  higher resolving power \citep[e.g.][]{Slob2024,Shajib2025}, which results in velocity resolution of $\leq$100 km\,s$^{-1}$ for median-resolution spectroscopy at the wavelength of \Ha. We define any emission-line with signal-to-noise (S/N) greater than three as detected, placing 3$\sigma$ limits derived from the \texttt{msaexp} noise spectrum, on non-detections. 

To understand whether the emission-line properties are originating from star-forming regions of the galaxy or in fact driven by Type-1 Active Galactic Nuclei (AGN) activity, we further fit each emission line feature as a double Gaussian with a narrow and broad component\footnote{Although, we note that broad emission-line features can also originate outflows in the galaxies which can be driven by star-formation or AGN activity \citep[e.g.][]{Fluetsch2021}.}. We constrain the fitting such that the FWHM of the narrow component is less than 800\,km/s and has independent line ratios whilst having consistent redshift. To evaluate the presence of the broad component we calculate the reduced chi-squared ($\chi^{2}_{\rm red.}$) and Bayesian Information Criteria (BIC) of each model, defined as,
\begin{align}\label{BIC}
    \chi^{2}\,&= \, \sum_{i}^{N_{\rm d}} r_{i}^{2} \\
    \chi^{2}_{\rm red.}\,&= \, \chi^{2} / (N_{\rm  d}\,- \,N_{\rm  varys})\\
    {\rm BIC}\,&=\, \chi^{2}+N_{\rm  varys} \times \ln(N_{\rm d})
\end{align}
where $\displaystyle\sum^{N_{\rm d}}_{i}r_{i}^{2}$ is the sum of the residual spectrum, $N_{\rm d}$ is the number of data points, and $N_{\rm  varys}$ is the number of variable parameters. Both of these statistical parameters reflect the goodness of fit whilst the BIC parameter penalises models with a large number of parameters and is commonly used to determine between two parametric models \citep[e.g.][]{Head2014,Lange2016}. 
\tcr{We require a $\Delta$BIC\,$<$$-$10 to statistically confirm the presence of a broad component in the spectrum of the sources}.

\subsection{Final sample} \label{Sec:Final}

Having constrained the photometric and spectroscopic properties of the 121 ALMA sources detected with NIRSpec grating spectra, we now define a final sample with robust properties. First we remove 14 sources for which \texttt{msaexp} can not determine the spectroscopic redshift, due to low signal to noise and or lack of emission lines. Next, we remove 52 sources where, given the spectroscopic redshift of the source, the rest-frame wavelength of the NIRSpec spectra does not encapsulate the \NH\ complex (\tcr{44 sources}), which is crucial to measure the  gas-phase metallicity, or for which have \Ha\ is not detected (S/N$<$3) (\tcr{8 sources}\footnote{\tcr{These eight sources exhibit similar flux densities and redshifts to the final sample of \FIRS\ galaxies, suggesting higher levels of dust obscuration is not the cause of non detection.}}). We ensure all sources have a well sampled observed-frame optical to near-infrared (0.4\,--\,1.6\um)  with a minimum of four filters and we thus remove three sources that do not have JWST (NIRCam or MIRI) imaging. Similarly, in the far-infrared we remove four sources whose ALMA detection is less than 3$\sigma$.

This results in a final sample of 48 sources with $>$3$\sigma$ ALMA continuum and \Ha\ detections, as presented in Table \ref{App:Selection}. The final sample has a median spectroscopic redshift of $z$\,=\,2.535\,$\pm$\,0.100 with a 16\qth\,--\,84\qth percentile range of $z$\,=\,1.830\,--\,3.855. We further derive a median stellar mass $\log_{10}$(M$_{\rm \ast}$[M$_{\odot}$])\,=\,10.8\,$\pm$\,0.1 with a 16\,--\,84th percentile range of $\log_{10}$(M$_{\rm \ast}$[M$_{\odot}$])\,=\,10.5\,--\,11.2, whilst from the \texttt{mercurius} modelling we establish a median dust mass of $\log_{10}$(M$_{\rm d}$[M$_{\odot}$])\,=\,8.7\,$\pm$\,0.1 with a 16\,--\,84th percentile range of $\log_{10}$(M$_{\rm d}$[M$_{\odot}$])\,=\,8.1\,--\,9.1. A summary of the SED derived properties for the final sample is given in Appendix \ref{App: SED_table}. Having established a robust sample of 48 ALMA sources with well-defined photometric and spectroscopic properties we now investigate their ionised interstellar medium properties in the context of the broader galaxy population at a similar epoch.

\section{Results and discussion}\label{Sec:Results}

Due to the inhomogeneous nature  of the far-infrared selection, our final sample of ALMA sources covers a broad range sub-millimetre flux density. As detailed in Table \ref{App:Selection}, 22 sources (46\%) are detected in ALMA Band 7, of which 9 are part of the ALMA SCUBA-2 UDS survey \citep[AS2UDS;][]{Stach2019,Dud2020} and represent sub-millimetre bright ($>$1\,mJy) galaxies (SMGs) with 850\um\ flux densities ranging from $S_{\rm B7}$\,=\,1.02\,--\,7.70\,mJy  which is comparable to the full AS2UDS sample \citep[e.g.][]{Dud2020} and the broader SMG population at this epoch \citep[e.g.][]{Hodge2013,Casey2014,Simpson2020}. The rest of the sample is made up of ALMA Band 6 (35\%, \tcr{17 sources}), Band 4 (15\%, \tcr{7 sources}), Band 8 (2\%, \tcr{1 source}) and Band 9 (2\%, \tcr{1 source}) with a median ALMA signal to noise of S/N\,=\,5.72\,$\pm$\,0.47 and a 16\qth\,--\,84\qth percentile range of S/N\,=\,3.53\,--\,9.27.

To homogenise  the sample, we utilise the far-infrared luminosity (L$_{\rm FIR}$), as derived from blackbody fitting (see Sec \ref{Sec:BB}). We note the L$_{\rm FIR}$ as derived from \texttt{mercurius} corresponds to the integrated blackbody emission and does not include warm or mid-infrared dust components. Typically the infrared luminosity (8\um\,--\,1000\um) is a factor of $\approx$1.6$\times$ ($\sim$0.2\,dex) higher than the far-infrared luminosity (40\um\,--\,1000\um) \cite[e.g.][]{Remy2015,Dud2020}. The \FIRS\ sample has a median far-infrared luminosity of log$_{10}$(L$_{\rm FIR}$[L$_{\odot}$])\,=\,11.95\,$\pm$\,0.08 with a 16\qth\,--\,84\qth percentile range of log$_{10}$(L$_{\rm FIR}$[L$_{\odot}$])\,=\,11.47\,--\,12.38. 

By construction, the sample encapsulates a full range of far-infrared luminosities from far-infrared bright systems comparable to those from AS2UDS with a median value of log$_{10}$(L$_{\rm IR}$[L$_{\odot}$])\,=\,12.45\,$\pm$\,0.02 \citep{Dud2020, Gillman2024} or the ALMA LABOCA ECDFS Submillimeter Survey \citep[ALESS;][]{Hodge2013,Swinbank2014} sample with a median value of log$_{10}$(L$_{\rm IR}$[L$_{\odot}$])\,=\,12.47\,$\pm$\,0.04 at a median redshift of  $z$\,=\,2.5\,$\pm$\,0.2. The sample also includes nine lower luminosity (log$_{10}$(L$_{\rm FIR}$[L$_{\odot}$])\,$<$\,11.5) sources, in the regime of luminous infrared galaxies \citep[LIRGS;][]{Sanders1988} comparable to studies of fainter DSFGs \citep[e.g.][]{Elbaz2011, Aravena2016,Tomoko2021,Mckay2025,Kiyota2026}.

\begin{figure*}[!htbp]
  \centering
  \begin{subfigure}{0.48\textwidth}
    \includegraphics[trim=0cm 0cm 32cm 0cm, clip, width=\linewidth]{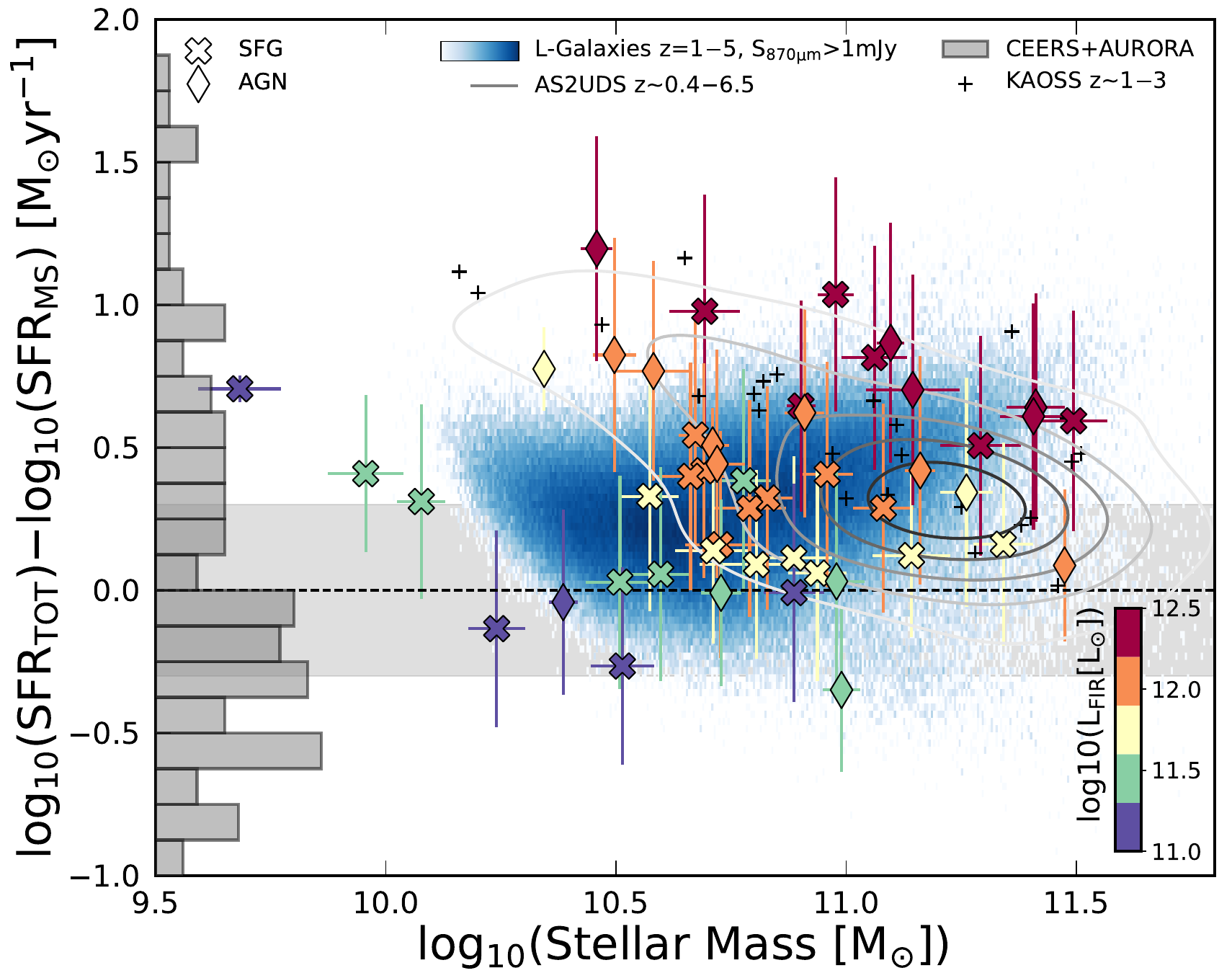}
  \end{subfigure}
  \begin{subfigure}{0.505\textwidth}
    \includegraphics[trim=30.5cm 0cm 0cm 0cm, clip, width=\linewidth]{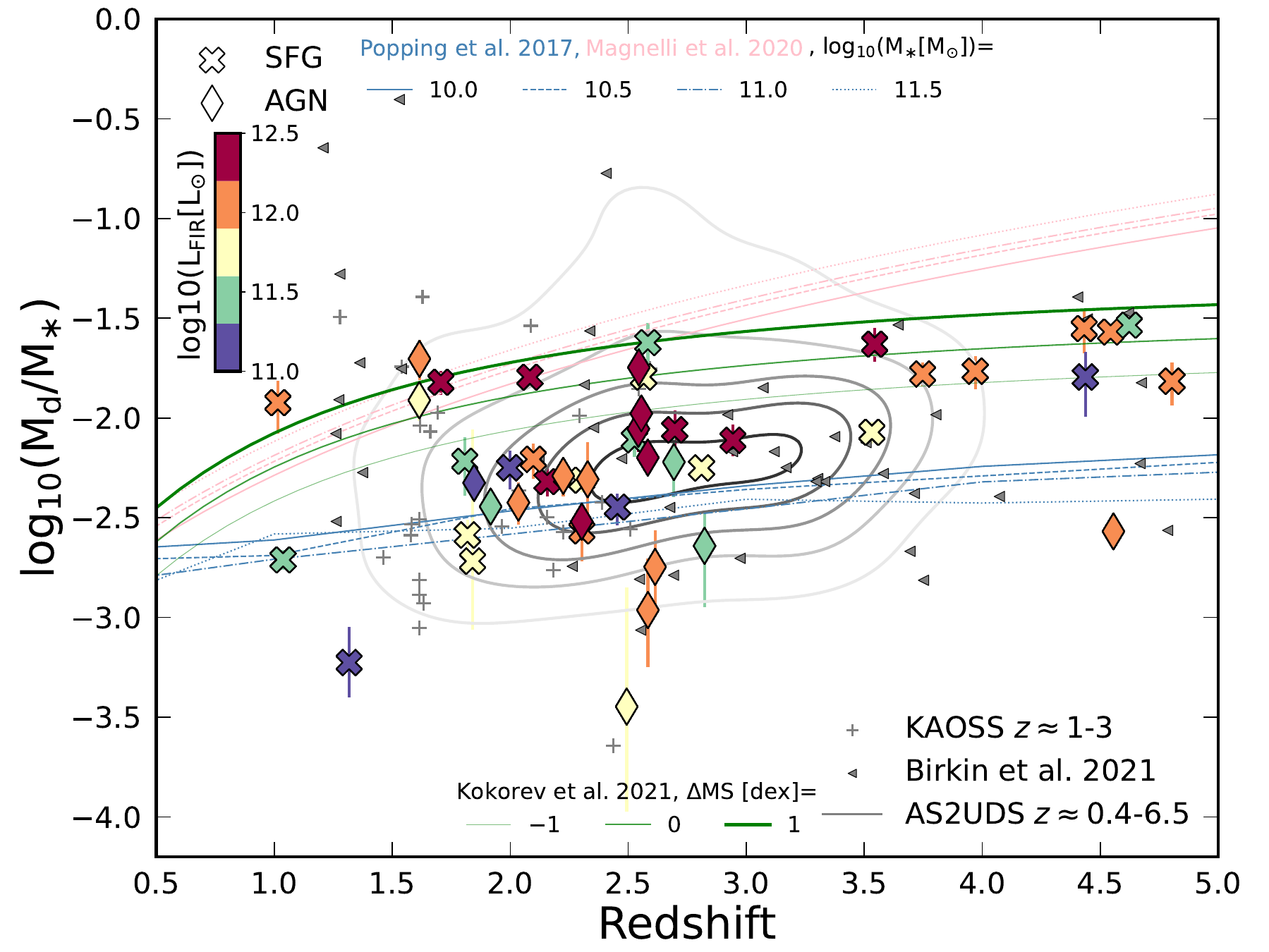}
  \end{subfigure}
  \caption{{Left}: The offset to the star-forming main sequence  relation from \citet{Popesso2023} as function of stellar mass. The ALMA-NIRSpec galaxies are coloured by their far-infrared luminosity (L$_{\rm FIR}$) \tcr{with crosses indicating potential AGN (Sec. \ref{Sec:AGN}), whilst sources with no signs of AGN activity (SFGs) are indicated by diamonds}. We further indicate the intrinsic 0.3\,dex scatter in the main-sequence (grey shaded region). We include comparison samples of high-redshift star-forming galaxies from the CEERS and AURORA surveys (grey histogram), as well as \Ha-detected SMGs from the KAOSS survey \citep{Birkin2022, Birkin2024} (grey plusses) and AS2UDS survey \citep{Dud2020} (grey contours), in addition to far-infrared bright ($S_{\rm 870\mu m}$\,$>$\,1mJy) galaxies at $z$\,=\,1\,--\,5 (blue density region) from the L-Galaxies SAM \citep{Araya2025}. 
  The ALMA-NIRSpec sample shows a strong correlation between main sequence offset and far-infrared luminosity, with far-infrared bright log$_{10}$(L$_{\rm IR}$[L$_{\odot}$])\,$>$\,12.0 galaxies falling ($\gtrsim$0.5\,dex) above the main-sequence relation, consistent with other bright far-infrared detected, and simulated, galaxies at the same epoch. 
    {Right}: The dust to stellar mass ratio as a function of redshift with the \FIRS\ galaxies coloured by far-infrared luminosity. We again include comparison samples of high-redshift SMGs from KAOSS, AS2UDS as well as CO-selected SMGs from \citet{Birkin2021}.
    Our sample has similar dust to stellar mass ratios when compared to traditional SMG samples at the same redshift. We also show the observational scaling relation from \citet{Magnelli2020} (pink lines) and the    
    Santa-Cruz SAM (blue lines) \citep{Popping2017} for a range of stellar masses \tcr{(solid, dashed, dot-dashed, dotted; $\log_{10}$(M$_{\rm \ast}$[M$_{\odot}$])\,=\,10.0, 10.5, 11.0, 11.5 respectively)} in addition to the observationally-derived scaling from \citet{Kokorev2021} at fixed stellar mass ($\log_{10}$(M$_{\rm \ast}$[M$_{\odot}$])\,=\,10.8) as a function of main sequence offset (green lines).
    }
 \label{fig:dMS_Md}
\end{figure*}

In Figure \ref{fig:RGB} we show the NIRCam false colour images for the 48 \FIRS\ galaxies, centred on the NIRCam F444W detected source, ranked in descending far-infrared luminosity. For each source we select the NIRCam filters that sample the rest-frame 2\um\ (R), 0.7\um\ (G) and 0.3\um\ (B). We further indicate the NIRSpec-ID, spectroscopic redshift, far-infrared luminosity and ALMA source position. From Figure \ref{fig:RGB} it is clear the sample represents a broad range of optical to near-infrared morphologies from isolated discs and compact point sources (e.g. ID:175747, ID:175918) to extended, potentially interacting sources (e.g. ID:173952, ID:172382). Their is no clear morphological correlation with far-infrared luminosity, with both high luminosity (log$_{10}$(L$_{\rm IR}$[L$_{\odot}$])\,$>$\,12.0) and lower luminosity (log$_{10}$(L$_{\rm IR}$[L$_{\odot}$])\,$<$\,11.5) galaxies, exhibiting a range of rest-frame UV/optical/infrared colours and morphologies. This is also in agreement with recent morphological studies of far-infrared detected galaxies \citep[e.g.][]{Gillman2024,Polletta2024,Chan2025}.

In terms of rest-frame optical morphology the sample has a median 0.5\um\ size of $R_{\rm  e,\,0.5\um}$\,=\,2.74\,$\pm$\,0.33\,kpc. Adopting rest-frame optical mass size relation from \citet{Ward2024_size} which is derived for star-forming galaxies in the CEERS survey, we establish on average the ALMA sources fall slightly below the relation with $\Delta R_{\rm  e,\,0.5\um}$[dex]\,=\,log$_{\rm 10}$($R_{\rm e,\,0.5\um, PySersic}$)\,$-$\,log$_{\rm 10}$($R_{\rm e,\,0.5\um, Ward}$)\,=\allowdisplaybreaks\, $-$0.08\,$\pm$\,0.03, although we note the \citet{Ward2024_size} relation is defined at lower ($<$10$^{10.5}$M$_{\odot}$) stellar masses. Splitting the sample into far-infrared bright sources (log$_{10}$(L$_{\rm IR}$[L$_{\odot}$])\,$>$\,12.0) and far-infrared faint sources (log$_{10}$(L$_{\rm IR}$[L$_{\odot}$])\,$<$\,11.5), we establish no difference in optical sizes at fixed stellar mass, with $\Delta R_{\rm  e,\,0.5\um}$[dex]\,=\,$-$0.11\,$\pm$\,0.03 and $\Delta R_{\rm  e,\,0.5\um}$[dex]\,=\,$-$0.11\,$\pm$\,0.06, respectively. This is consistent with recent studies of far-infrared detected galaxies with JWST, which suggest DSFGs follow the same stellar mass size relation as less dust obscured sources \citep[e.g.][]{Boogaard2024,Gillman2024, Mckinney2024,Ren2024}. \tcr{We note that heavily obscured central regions of the galaxies would result in more extended observed optical profiles in comparison to the near-infrared stellar continuum, that is less affected by dust. This affect was observed in \citet{Gillman2024}, with 870\um\ selected sources displaying 50\% larger sizes in the rest-frame $V$-band ($\sim$0.5\um), compared to rest-frame $z$-band ($\sim$1\um), which fall slightly below the rest-frame optical mass-size relation (see \citet{Gillman2024} for details.)}

\subsection{Star-formation and dust content}

In Figure \ref{fig:dMS_Md}a, we show the position of the ALMA-NIRSpec galaxies in relation to the star-forming main-sequence at their epoch as defined from \citet{Popesso2023}\footnote{We not the exact offset depends on the functional form of the main sequence, where a liner power-law \citep[e.g.][]{Pearson2018} would result in a smaller offset at the highest masses.}. For each galaxy we define the total star-formation rate (SFR$_{\rm tot}$) as the observed (not dust corrected) H$\alpha$ star-formation rate (SFR$_{\rm H \alpha}$)\footnote{We note here we are assuming all \Ha\ narrow flux is due to star formation activity, so in this respect is an upper limit.} plus SFR$_{\rm FIR}$ as derived in Sec. \ref{Sec:BB}. \tcr{We adopt this approach given the total star formation rate in the \FIRS\ galaxies is dominated by the far-infrared component, and it further minimises assumptions regarding dust attenuation prescriptions  as is assumed in the near-infrared to optical SED derived SFR.}  We further colour the \FIRS\ galaxies by their far-infrared luminosity, identifying a clear trend, of galaxies above log$_{10}$(M$_{\ast}$[M$_{\odot}$)\,$>$\,10.5 with log$_{10}$(L$_{\rm FIR}$[L$_{\odot}$])\,$<$\,11.5 falling on or below the main sequence, whilst those at log$_{10}$(L$_{\rm FIR}$[L$_{\odot}$])\,$>$\,11.5 lie above the main sequence at their stellar mass.

For comparison, we include sources from recent JWST NIRSpec surveys of typical star-forming galaxies where there is no (detectable) far-infrared star formation contribution. In particular we utilise results from the Assembly of Ultradeep Rest-optical Observations Revealing Astrophysics (AURORA) survey \citep{Shapley2025} and the Cosmic Evolution Early Release Science (CEERS) Survey \citep{Shapley2023}. From these surveys we selected a redshift matched sample of 112 (57 AURORA, 55 CEERS) star-forming galaxies that have no indication of broad-line AGN activity\footnote{\texttt{agnflag=0}, see \citet{Shapley2023,Shapley2025} for details.} and emission-lines fluxes at S/N$>$3. This subsample of CEERS+AURORA galaxies, as indicated by the grey histogram in Figure  \ref{fig:dMS_Md}a, has a median stellar mass of log$_{10}$(M$_{\ast}$M$_{\odot}$])\,=\,9.5\,$\pm$\,0.1 and a median SED derived star formation rate of SFR\,=\,6\,$\pm$\,2\,M$_{\odot}$\,yr$^{-1}$.

We further include literature samples of SMGs from the AS2UDS survey \citep{Dud2020} and the KMOS-ALMA Observations of Submillimetre Sources (KAOSS; \citealp[][]{Birkin2022,Birkin2024},Taylor et al. in prep.). For the  KAOSS sources we use the same SFR$_{\rm tot}$ definition as for the \FIRS\ sample (uncorrected \Ha\ star-formation rate plus far-infrared star-formation) as derived in \citet{Birkin2022}. We scale the \texttt{magphys} derived total star-formation rates of the AS2UDS sample by a factor of $\approx$1.3$\times$ which accounts for the scaling between  SFR$_{\rm tot}$, as defined above, and SFR$_{\rm magphys}$ (see \cite{Dud2020, Birkin2022} for details). In addition to the observational studies, we make use of recent results from L-Galaxies semi-analytic model \citep{Araya2025,Araya2025a} that is re-calibrated using an Markov chain Monte Carlo (MCMC) approach, to a provide an improved match to observationally-inferred number densities of both SMGs and massive quiescent galaxies populations. We select a sub-sample of far-infrared bright ($S_{\rm 870\mu m}$\,$>$\,1mJy) at $z$\,=\,1\,--\,5 galaxies, whose stellar mass and star-formation properties reflect the full  far-infrared bright population in L-Galaxies. We calculate the main-sequence offset of the L-Galaxies sample using
the instantaneous star-formation rate within the snapshot, which is equates to a 5.6\,Myr timescale at $z$\,=\,1 and increases to 18.19\,Myr at $z$\,=\,5\footnote{We note this is not exactly the same as the observational total star formation rate. The results of our analysis are unchanged if we adopt the snapshot average star formation rate from the simulation, which varies from 112\,Myr timescale at $z$\,=\,1 to 368\,Myr at $z$\,=\,5.}. 

On average the \FIRS\ galaxies fall 0.32\,$\pm$\,0.07\,dex above the main sequence, with the highest infrared luminosity sources lying well above the main sequence ($\gtrsim$\,0.5\,dex), with a similar main sequence offset to the AS2UDS and KAOSS samples at the same stellar mass. In contrast, the CEERS+AURORA sample (as indicated by the grey histogram in Figure \ref{fig:dMS_Md}a), of less-massive, less dust-obscured galaxies, scatter about the main-sequence with a median main sequence offset of $-$0.11\,$\pm$\,0.07\,dex, which is within the intrinsic $\approx$0.3\,dex intrinsic scatter of the main sequence relation \citep[e.g.][]{Popesso2023}. The L-Galaxies sources, with a 16\qth\,--\,84\qth stellar mass range of log$_{10}$(M$_{\ast}$M$_{\odot}$])\,=\,10.50\,--\,11.0, have a median main sequence offset of 0.28\,dex, which is just within the 0.3\,dex intrinsic scatter of the main-sequence, but predominately at higher specific star-formation rates compared to typical star-forming galaxies from CEERS and AURORA at the same epoch. For L-Galaxies sources above log$_{10}$(M$_{\ast}[$M$_{\odot}$])\,=\,11.0, we identify an increased main sequence offset of 0.43\,dex, in a similar trend to the \FIRS\ galaxies.

In Figure \ref{fig:dMS_Md}b, we show the dust to stellar mass ratio of the \FIRS\ galaxies as a function of redshift, coloured by their far-infrared luminosity. We again include the samples of SMGs from AS2UDS and KAOSS, as well as CO-selected SMGs from \citet{Birkin2021}. We rescale the dust masses of the AS2UDS and KAOSS samples, derived from \texttt{magphys}, by a factor 1.2 (as derived in \citealt{Dud2020}), to account for the differences between modified blackbody and full multi-wavelength SED derived dust masses (see also \citealt{Hunt2019}). The \FIRS\ galaxies have median dust fraction of log$_{10}$(M$_{\rm d}$/M$_{\rm \ast}$)\,=\,$-$2.20\,$\pm$\,0.08 with a 16\qth\,--\,84\qth\ range of  log$_{10}$(M$_{\rm d}$/M$_{\rm \ast}$)\,=\,$-$2.58 to $-$1.78. On average the sample is consistent with the sub-millimetre bright comparison samples at a median redshift of $z$\,=\,2.5, but exhibits higher dust fractions at higher redshift, with the \FIRS\ galaxies at $z$\,$>$\,3 having a median dust fraction of log$_{10}$(M$_{\rm d}$/M$_{\rm \ast}$)\,=\,$-$1.78\,$\pm$\,0.09.

In Figure \ref{fig:dMS_Md}b we include predictions from the fiducial model of the Santa-Cruz SAM \citet{Popping2014, Popping2017}, which tracks the lifecycle of dust with prescriptions for dust production, growth and destruction. We show the models predicted evolution of the dust fraction as a function of redshift (blue lines) for fixed stellar masses (log$_{10}$(M$_{\rm \ast}$[M$_{\odot}$])\,=\,10.0, 10.5, 11.0, 11.5). We also show the observationally-derived scaling relation from \citet{Magnelli2020}, for the same stellar mass ranges (pink lines), as well as the relation from \citet{Kokorev2021}, at the median stellar mass of the \FIRS\ sample (log$_{10}$(M$_{\rm \ast}$[M$_{\odot}$])\,=\,10.8), as a function of main sequence offset (green lines), at $-$1, 0 and 1\,dex off the main sequence. At the median redshift of the sample ($z$\,=\,2.5), the \FIRS\ sample is more aligned with the \citet{Popping2017} model, whose dust growth its implicitly tied to the metal budget of the galaxies, which is systematically lower than the dust fractions derived from far-infrared observations of main-sequence galaxies in \citet{Magnelli2020,Kokorev2021}. At higher redshift ($z$\,$>$\,3) the \FIRS\ sample, shows higher dust fractions, and is more aligned with the observationally derived \citet{Magnelli2020, Kokorev2021} relations.  

Figures \ref{fig:RGB} \& \ref{fig:dMS_Md} demonstrate that despite the inhomogeneous selection of our sample, on average the \FIRS\ galaxies reflect the broader population of uniformly selected (e.g. 870\um\ selected) DSFGs at their epoch, both in terms of morphology, and photometric properties (e.g. stellar mass, star formation rate and, dust mass). We now focus on analysing the properties of their ionised interstellar medium, as unprecedentedly revealed by the NIRSpec grating spectra.

\begin{figure*}[!htbp]
    \centering
    \includegraphics[width=\linewidth]{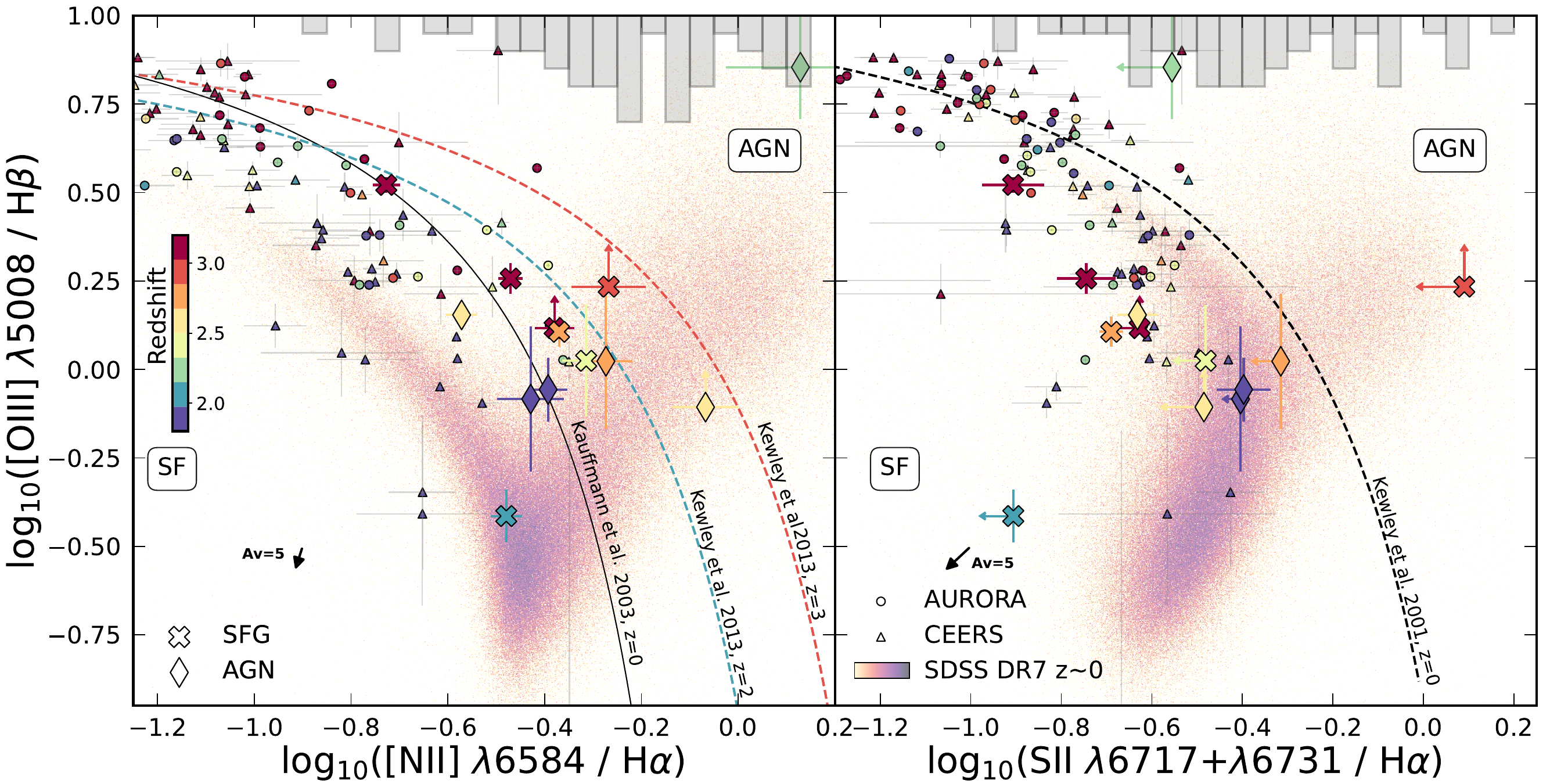}
    \caption{Observed (dust-uncorrected) \OTHb\ as a function of \NH\ (left) and \STHa\ (right) \tcr{with diamonds indicating AGN and crosses SFGs from the \FIRS\ sample}. Sources with S/N$<$3 in the respective emission-line are shown as limits. The  distribution of \NH\ and \STHa\ for the full sample is shown by the grey histograms. We show comparison samples of typical star-forming galaxies at $z$\,$\sim$\,0 (red density) from the Sloan Digital Sky Survey \citep[SDSS;][]{Bricnhmann2004}, as well as a redshift matched sample of star-forming galaxies from the CEERS \citep{Shapley2023} (\tcr{small triangles}) and AURORA Surveys \citet{Shapley2025} (\tcr{small circles}). The black solid curve shows the division between star-formation acticity  and AGN \citep[][]{Kauffmann2003} at $z$\,=\,0 whilst the theoretical demarcation at $z$\,=\,2 and $z$\,=\,3 from  \citep[][]{Kewley2013} is given by the blue- and red-dashed lines respectively. The black arrow indicates the impact of \av\,=\,5 dust correction using a \citet{Calzetti2000} dust law. The majority of \FIRS\ galaxies are consistent with typical SFGs at the same epoch, pushing to lower \OTHb\ line ratios, with only three galaxies showing AGN-like optical line diagnostics.}
    \label{fig:BPT}
\end{figure*}


\subsection{Ionisation origin}\label{Sec:AGN}

Prior to analysing the properties of the ionised interstellar medium of the \FIRS, we first investigate whether their emission is the result of AGN activity or star-formation.

First, we check for the X-ray counterparts to the \FIRS\ galaxies, requiring any source with an X-ray detection to be greater 10$^{42}$ erg/s to be flagged as an AGN. We adopt this threshold given that fainter X-ray emission can have stellar origins (e.g. X-ray binaries, shocks) and thus does not necessarily indicate the presence of AGN activity \cite[e.g.][]{Lehmer2010,Hickox2018}. 

For sources in the  UDS field we adopt the 600ks Chandra Legacy Survey of UDS as presented in \citet{Kocevski2018}, whilst for GOODS-S and COSMOS we adopt the 7Ms Chandra Deep Field-South survey as presented in \citet{Luo2017} and the 4.6Ms Chandra COSMOS-Legacy survey from \citet{Civano2016} respectively. We crossmatch our NIRSpec source positions to within 1\farcs{0} of the X-ray sources, which accounts for both the resolution and astrometry offsets between Chandra and JWST \citep[e.g.][]{Lyu2022, Gillman2025}. We identify 15 counterparts (31\% of the sample) within 0\farcs{8}, \tcr{all of which have an absorption corrected X-ray luminosity $>$10$^{42}$ erg/s}  with a median value of log$_{10}$(L$_{\rm Xc}$[erg/s])\,=\,42.73\,$\pm$\,0.26\,erg/s. Thirteen of the sources are from the \citet{Luo2017} catalogue in GOODS-S where the X-ray luminosity comes from the full (soft + hard) bands (0.5\,--\,7\,Kev). 
The two remaining sources are from the \citet{Kocevski2018} catalogue in UDS for which the X-ray luminosity represents the hard band (2\,--\,10\,kev). Whilst X-ray detections provide one way of identifying AGN activity in the \FIRS\ galaxies, as demonstrated by many high-redshift AGN studies  \citep[e.g.][]{Lyu2022, Lyu2024}, it is crucial to use multi-wavelength tracers that capture the full multi-phase emission of AGN that can originate from both heavily dust obscured  and unobscured regions \citep[e.g.][]{Padovani2017}. 

To further evaluate the presence of AGN activity in the \FIRS\ sample, we utilise  the optical emission line diagnostics. In particular we calculate the \OTHb, \STHa, and \NH\ emission lines ratios\footnote{\tcr{Line ratios are calculated for the narrow component of the emission line if a broad component is present (see Section \ref{Sec:Fitting}).}}, for which elevated line ratios can indicate the presence of harder ionizing spectrums and thus the presence of AGN activity \citep[e.g.][]{Kewley2001, Kauffmann2003} although as noted by many studies, high-metallicity star-bursting galaxies or low-metallicity AGN blur this demarcation  \citep[e.g.][]{Gutkin2016,Strom2017,Garg2022,Agostino2023,Hirschmann2023}.

\begin{figure*}[!htbp]
    \centering
    \includegraphics[width=\linewidth]{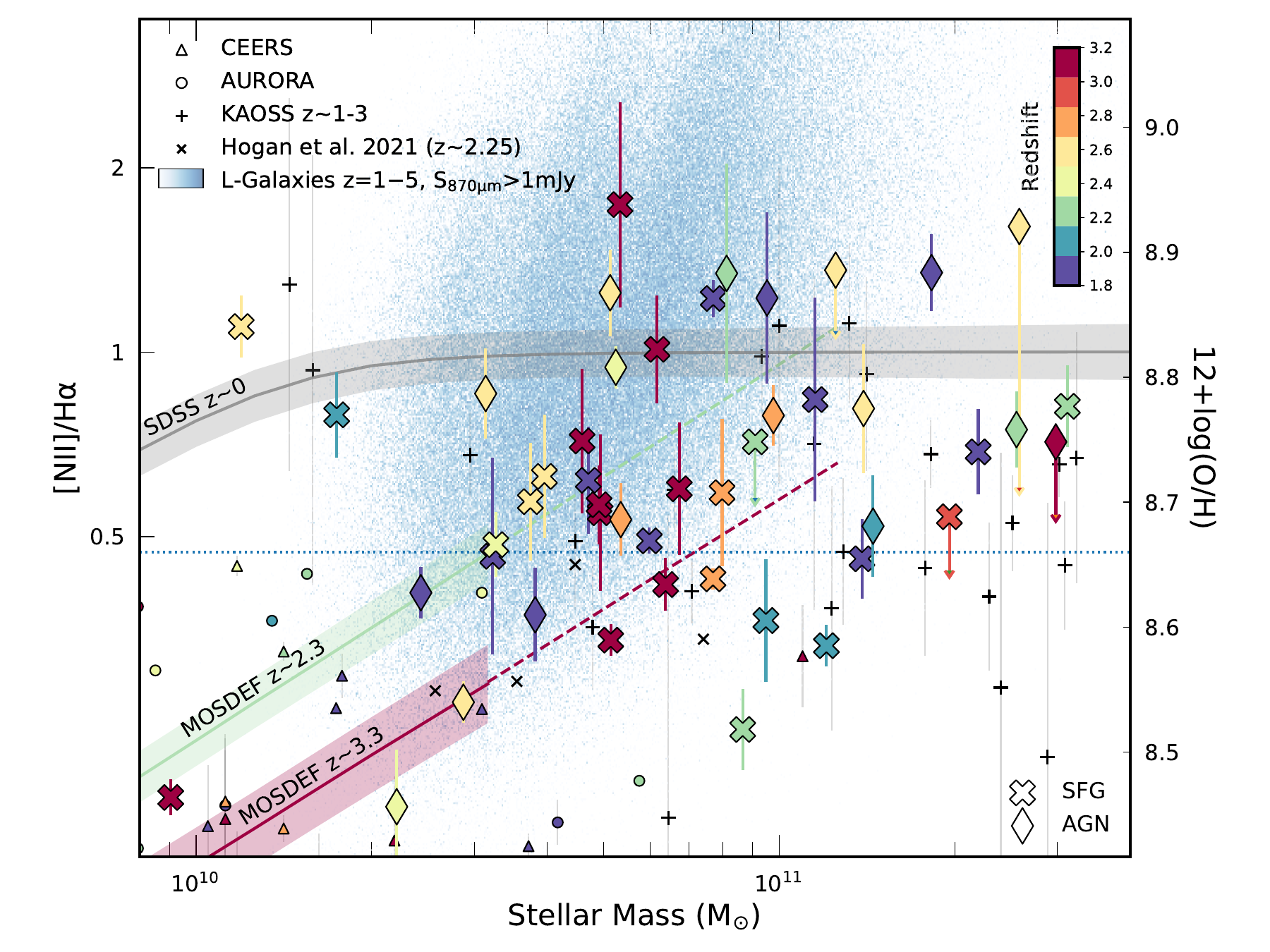}
    \caption{Gas-phase metallicity, as derived from the \NH\ index using the \citet{Bian2018} strong-line calibrations, as a function of stellar mass for the \FIRS\ galaxies \tcr{with diamonds indicating AGN and crosses SFGs}. Sources with S/N$<$3 in then \nii\ 6584$\lambda$ emission-line are shown as limits. The horizontal dashed line indicates solar metallicity \citep[12+log(O/H)= 8.66]{Asplund2005}.  We show comparison samples of typical star-forming galaxies at $z$\,$\sim$\,0 from SDSS \citep{Curti2020}  (grey line), the MOSDEF Survey \citep{Sanders2021} at  $z$\,$\sim$\,2.3 (green solid line) and  $z$\,$\sim$\,3.3 (red solid line), and a redshift matched sample of galaxies from the CEERS \citep{Shapley2023} (\tcr{small triangles})  and AURORA Surveys \citet{Shapley2025} (\tcr{small circles}) . We further compare to high-redshift observations of DSFGs from KAOSS at $z$\,$\sim$\,1\,--\,3  \citep{Birkin2022, Birkin2024} and Herschel selected sources from \citet{Hogan2021} at $z$\,$\sim$\,2.25, as well as far-infrared bright ($S_{\rm 870\mu m}$\,$>$\,1mJy) at $z$\,=\,1\,--\,5 galaxies (blue density) from the L-Galaxies simulation \citep{Araya2025}. The \FIRS\ galaxies have higher gas-phase metallicity than the observational comparison samples at the same epoch, but comparable to predications from L-Galaxies.}
    \label{fig:MZR}
\end{figure*}


For 14 of the \FIRS\ galaxies we have detections (S/N$>$3) of the \oiii\,5008$\lambda$ emission line of which 10 sources have further detections of the \Hb\ emission-line (see Appendix \ref{App:NIRSpec_table}). For the four sources without \Hb\ detections, we define 3$\sigma$ upper limits on the \OTHb\ line ratio if the NIRSpec spectra of the source covers the \Hb\ emission line.  In Figure \ref{fig:BPT}, we construct the Baldwin, Phillips \& Terlevich (BPT) diagram \citep{Baldwin1981}, by correlating the observed (not dust-corrected) \OTHb\ emission-line ratio with the \NH\ and \STHa\ emission-line ratios. As indicated  by the black arrow in Figure \ref{fig:BPT}, which depicts the magnitude of a dust correction using a \citet{Calzetti2000} dust law with \av\,=\,5, the impact of the dust-correction on the line ratios is minimal given their spectral proximity. We note however the line ratios shown are slight upper limits as we do not account for stellar absorption of the Balmer lines, which has been shown to result in $\sim$2\% emission-line flux reduction in post starburst galaxies with young stellar populations \citep[e.g.][]{Delgado1999,Goto2003,Reddy2018}.

In Figure \ref{fig:BPT}, we further include the redshift matched comparison sample of star-forming galaxies from the CEERS and AURORA surveys. Compared to the CEERS and AURORA comparison samples at the same epoch, the \FIRS\ galaxies exhibit lower \OTHb\  ratios with a median value of log$_{10}$(\OTHb)\,=\,0.11\,$\pm$\,0.08 more comparable to local galaxies, which can potentially indicate a metal rich interstellar medium with low ionisation parameters \citep[e.g.][]{Steidel2014,Strom2017,Kashino2019}. In Figure \ref{fig:BPT}, we show the demarcation lines between SF and AGN from \citet{Kauffmann2003} at $z$\,=\,0 as well as the redshift evolving division from \citet{Kewley2013} at $z$\,=\,2\,--\,3 \tcr{as given by,
\begin{equation}
\log\left(\frac{[\mathrm{O\,III}]}{\mathrm{H}\beta}\right)
=1.1 + 0.03z+\frac{0.61}{\log\left(\frac{[\mathrm{N\,II}]}{\mathrm{H}\alpha}\right) + 0.08 - 0.1833z}
\end{equation}}
\tcr{whereby at higher redshift the division between SF and AGN occurs at higher values of \NH\ and \OTHb\ index}. By evaluating the demarcation at the redshift of each \FIRS\ galaxies we establish the majority (94\%) occupy the SF region at their respective epoch, with three galaxies exhibiting enhanced \OTHb\ ratios for their \NH\ line ratios, indicating their emission properties  do not to purely originate from star formation activity alone. All three of these galaxies are detected in the \citet{Luo2017} 7Ms Chandra catalogue of GOODS-S and have intrinsic X-ray luminosities greater than 10$^{42}$ erg/s.\footnote{\tcr{We note adopting the \citet{Kauffmann2003} relation at $z$\,=\,0 for all sources results in nine sources being classified as AGN from optical line diagnostics.}}

Given not all \FIRS\ galaxies have detections of the \OTHb\ line ratio, we also investigate the  distribution of \NH\ index in the sample, as shown by the grey histogram in  Figure \ref{fig:BPT}a. The \FIRS\ sample has a median value of \NH\,=\,0.61\,$\pm$\,0.06, which is below the typical threshold of \NH\,$\gtrsim$\,0.8  used to indicate AGN activity \citep[e.g.][]{Wisnioski2018,Leung2019,Birkin2023}, suggesting on average that their ionisation is not driven by AGN activity. \tcr{Of the 48 sources 15 (31\%) exhibit \NH\,$\gtrsim$\,0.8, which if adopted as a more conservative AGN criteria would result in an AGN fraction of 52\%.}

As well as optical emission-line ratio diagnostics, we can further use the profile of the rest-frame optical emission lines to identify the presence of broad-line (Type-1) AGN in the \FIRS\ sample. Using the BIC parameter (Equation \ref{BIC}), from the emission-line fitting routine (see Sec \ref{Sec:Fitting}), we can conduct a statistical analysis on the presence of broad line features. The \FIRS\ galaxies have a median \Ha\ FWHM of 328\,$\pm$\,27\,km/s\footnote{Excluding the 19 AGN from the sample, the median \Ha\ FWHM becomes 305\,$\pm$\,22\,km/s.}.  Using the threshold of $\Delta$BIC$_{\rm H\alpha,[\textsc{nii}],[\textsc{sii}]}$\,$<$$-$10 to indicate the presence of broad lines we identify nine galaxies (19\%) that fulfil this criteria, of which five have X-ray counterparts with luminosities greater than 10$^{42}$\,erg/s. For these nine sources, with a median \Ha\ broad component  FWHM of 1818\,$\pm$\,241\,km/s, we report the narrow-line component emission-line ratios in all of the forthcoming analysis. \tcr{For three of these sources there is no spectral coverage of the \OTHb\ complex, whilst four sources have none detections (S/N$<$3) of the \OTHb\ complex. The two remaining sources with broad \Ha\ lines have equally broad (FWHM$>$800\,km/s) \OTHb\ emission lines.}

In total, we identify 19 (40\%) of the \FIRS\ galaxies have indications of AGN activity either through X-ray counterparts, optical emission-line ratios or broad line components. Whilst we report the fraction of galaxies which show indications of AGN activity, we note given the sub-optimal placement of the NIRSpec slits for our sources (see Figure \ref{fig:RGB}), there might be sources for which MSA shutter misses the central, core, region of the galaxy, that may harbour an AGN, thus we note this is a lower limit on AGN activity. Throughput the paper we highlight these 19 sources as AGN.

\subsection{Gas-phase metallicity}

Having identified the potential AGN contributions to the \FIRS\ galaxies emission, we now investigate the metallicity properties of the sample. Given the lack of coverage and low signal to noise ($<$1$\sigma$) of the \oii\ $\lambda$4363 emission-line in the NIRSpec spectra, it is not possible to derive the gas-phase metallicity using the `direct-method' based on electron temperature \citep[e.g.][]{Pollock2025,Kostovan2025}. However,  we can use strong-line calibrations which give the relation between the higher signal-to-noise nebular lines and the gas-phase Oxygen abundance \citep[e.g.][]{Curti2020,Gillman2021,Gillman2022}. 

In particular we adopt the relation from \citet{Bian2018} for the \NH\ index for which we have detections in 90\% (43/48) of the sample. For the six remaining sources the \nii\,6584$\lambda$  emission-line has a S/N$<$3 so we adopt a 3$\sigma$ upper limit.
The \citet{Bian2018} calibration is derived from stacked spectra of nearby galaxies whose \OTHb\ and \NH\, emission-line ratios place them in the same region of parameter space in the BPT diagram as those at higher redshift ($z$\,$\approx$\,2). Specifically the gas-phase metallicity is given by,
\begin{equation}
    \label{Eq:Bian}
   12+\log_{10}(\text{O/H})\,=\,8.82+\log_{10}(\text{\NH})\,\times\,0.49)
\end{equation}
As reported in \citet{Sanders2021}, the \NH\ strong-line calibration of \citet{Bian2018} has an intrinsic $\approx$0.11\,dex scatter. For the \FIRS\ galaxies, including the upper limits, we derive a median \NH\, ratio of \NH\,=\,0.61\,$\pm$\,0.06 with a 16\,--\,84th percentile range of  \NH\,=\,0.37\,--\,1.16. This results in a median gas-phase metallicity of \ZOH\,=\,8.71\,$\pm$\,0.02, with a 16\,--\,84th percentile range of  \ZOH\,=\,8.61\,--\,8.85\footnote{Adopting the \citet{Pettini2004} strong-line calibrations results in a 0.16\,dex higher metallicity.}. To understand if the sub-optimal placement of NIRSpec slits is biasing our results, we measure the gas-phase metallicity in a subsample of 28 galaxies whose \FIRS\ separation is $<$0\farcs{1}, median separation of 0\farcs{05}, identifying a consistently high metallicity of \ZOH\,=\,8.76\,$\pm$\,0.03. \tcr{Finally, to ensure the metallicities are not driven by the presence of AGN in the sample, we calculate the median \NH\ index and gas-phase metallicity of 29 sources that show no signs of AGN activity. We derive a median \NH\,=\,0.57\,$\pm$\,0.14 and \ZOH\,=\,8.70\,$\pm$\,0.06,  consistent with the full \FIRS\ sample}\footnote{\tcr{We note if we adopt the more conservative \NH\,$\gtrsim$\,0.8 AGN criteria resulting in a 52\% AGN fraction (Sec. \ref{Sec:AGN}), we establish a consistent metallicity of \ZOH\,=\,8.68\,$\pm$\,0.04 for the non-AGN sources.}}.

\begin{figure*}[!htbp]
  \centering
  \begin{subfigure}{0.49\textwidth}
    \includegraphics[trim=29cm 0cm 2.5cm 0cm, clip, width=\linewidth]{Figures_pdf/dMS_FMR.pdf}
  \end{subfigure}
  \begin{subfigure}{0.5\textwidth}
    \includegraphics[trim=0.8cm 0cm 0cm 0cm, clip,width=\linewidth]{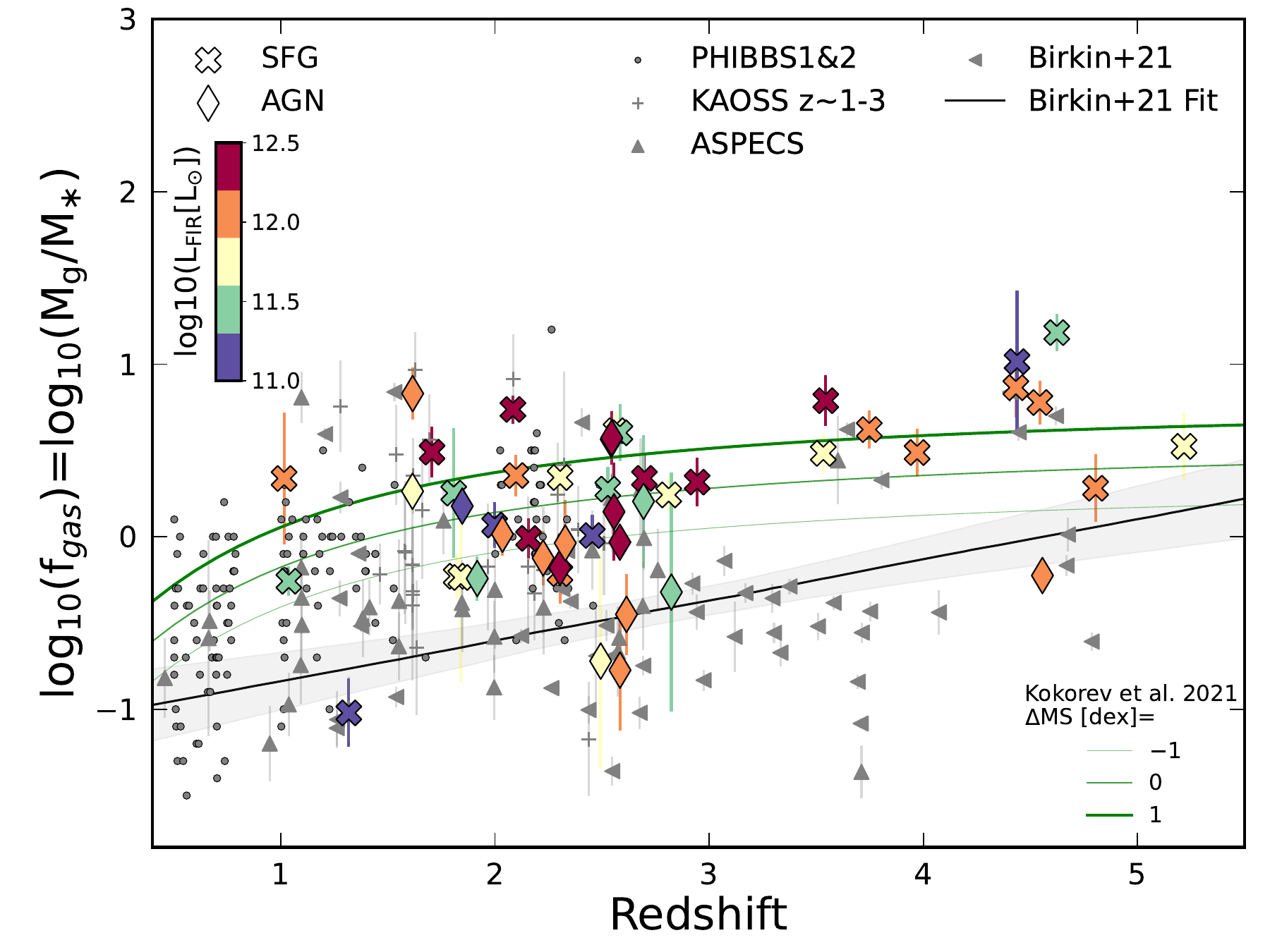}
  \end{subfigure}
  \caption{Left: The metallicity offset to the fundamental metallicity relation (FMR) as defined by \citet{Curti2020}, with the 0.05\,dex uncertainty on the relation (grey shaded region) \tcr{with diamonds indicating AGN and crosses SFGs from the \FIRS\ sample}. We include comparison samples from the CEERS (\tcr{black triangles}) and AURORA \tcr{(black circles)} surveys \citep{Shapley2023,Shapley2025} as well as sub-millimetre bright ($>$1\,mJy) galaxies in the L-Galaxies simulation (histograms) at $z$\,=\,1\,--\,5 \citep{Araya2025}, with the median at each redshift snapshot indicated by the black line. 
     On average the \FIRS\ galaxies show a 0.13\,$\pm$\,0.03\,dex elevation in metallicity compared to the FMR, which is slightly lower than the predictions from L-Galaxies. 
    Right: Gas Fraction (M$_{\rm g}$/M$_{\ast}$) as a function redshift for the  \FIRS\ galaxies coloured by their total specific star-formation rate \tcr{with diamonds indicating AGN and crosses SFGs from the \FIRS\ sample}. We include comparison samples of SFGs from the PHIBBS \citep{Tacconi2018} and ASPECS \citep{Boogaard2020} surveys, SMGs with CO-derived gas masses from \citet{Birkin2021} and SMGs with metallicity derived gas masses from KAOSS \citep{Birkin2022,Birkin2024}. We further over plot the best fitting relation from \citet{Birkin2021} (black line). The  \FIRS\ galaxies show elevated gas fractions compared to the CO-detected samples, more consistent with the main sequence scaling from \citet{Kokorev2021}.}
  \label{fig:dFMR_MG}
\end{figure*}

In Figure \ref{fig:MZR}, we plot the gas-phase metallicity, as a function of \texttt{bagpipes} derived stellar mass for the \FIRS\ galaxies, coloured by their spectroscopic redshift. We show the well established scaling relations derived for typical star-forming galaxies in the local Universe ($z$\,$\sim$\,0) from SDSS \citep{Curti2020}, and the MOSFIRE Deep Evolution Field \citep[MOSDEF:][]{Kriek2015} Survey at $z$\,$\sim$\,2.3 and  $z$\,$\sim$\,3.3 \citep{Sanders2021}, which on average are defined at lower stellar masses (log$_{10}$(M$_{\rm \ast}$[M$_{\odot}$])\,$<$\,10.6). We further include the redshift matched samples of typical star-forming galaxies from CEERS and AURORA as well as the SMGs from KAOSS (\citealp{Birkin2022}, Taylor et al. in prep.) and Herschel selected sources from \citet{Hogan2021} at $z$\,$\sim$\,2.25. Finally, we include the predictions for the sub-millimetre bright sources from the L-Galaxies SAM at $z$\,=\,1\,--\,5 \citep{Araya2025,Araya2025a}.

On average we identify high gas-phase metallicities in the \FIRS\ galaxies, compared to other SFGs at the same epoch, comparable to the metallicities at $z$\,$\sim$\,0. At the highest stellar masses, the sample is comparable to that reported in the KAOSS Survey \citep[12+log(O/H)\,$\simeq$\,8.75;][]{Birkin2022}, and in tentative agreement with extrapolations of the SFGs relations (dashed lines) defined at lower stellar masses. Below 10$^{11}$M$_{\odot}$, the \FIRS\ galaxies exhibit high metallicities in agreement with the predictions from L-Galaxies. The majority of L-Galaxies sample falls below a stellar mass of 10$^{11}$M$_{\odot}$, with a median stellar mass of log$_{10}$(M$_{\rm \ast}$[M$_{\odot}$])\,=\,10.73. As noted in \citet{Araya2025a}, whilst the semi-analytical model matches the number counts of SMGs, there is an offset in the galaxy stellar mass function, with the model lacking the highest mass galaxies. 

The super solar metallicities of the \FIRS\ galaxies is also in agreement with \citet{Kiyota2026} which identified 0.4\,--\,2\,$\times$ solar metallicity in 16 faint (median log$_{10}$(L$_{\rm IR}$[L$_{\odot}$])\,=\,11.62\,$\pm$\,0.14) ALMA detected sources from the ASPECS survey in GOODS-S with a median redshift and stellar mass of $z$\,=\,2.04\,$\pm$\,0.23 and log$_{10}$(M$_{\rm \ast}$[M$_{\odot}$])\,=\,10.31\,$\pm$\,0.20. This is also in agreement with previous studies of the gas-phase metallicity in SMGs using the fine structure lines from Herschel-PACS observations \citep{Wardlow2017} and targetted ALMA observations at higher redshift ($z$\,$>$\,4), indicating the presence of a chemically evolved interstellar medium \citep{Tadaki2019}. Although, we note this is somewhat higher than the 0.9\,$\times$ solar metallicity reported by \citet{Tomoko2021} for $z$\,$>$\,3 faint ALMA Band 6 ($S_{\rm B6}$\,=\,0.11\,$\pm$\,0.15\,mJy) detected galaxies, which was derived using the metallicity calibrations of \citet{Curti2017} and not based on the \NH\ index. %

\subsection{Fundamental metallicity relation}

To understand whether the \FIRS\ galaxies occupy the same fundamental metallicity relation (FMR) as typical SFGs, in Figure \ref{fig:dFMR_MG}a we plot the offset in terms of gas-phase metallicity between the measured metallicity and that predicted by the plane, as parametrised by \citet{Curti2020} (see their Equation 5), for a given stellar mass and total star-formation rate of the \FIRS\ galaxies. We measure the same metallicity offset for the CEERS, AURORA, KAOSS and L-Galaxies samples. On average the \FIRS\ galaxies have an 0.13\,$\pm$\,0.03\,dex higher metallicity than predicted by the FMR, whilst being lower than the L-Galaxies model which at $z$\,=\,2\,--\,3, has an average metallicity that is 0.22\,dex higher than the FMR. The SFGs from CEERS and AURORA, for which we use the \Ha\ (dust corrected)\footnote{Assuming a \citet{Calzetti2000} attenuation law and Case B recombination.} star-formation rate following the definition from \citet{Curti2020}, are consistent with the relation with a median offset of $-$0.03\,$\pm$\,0.02\,dex. \tcr{To ensure the offset of the \FIRS\ galxies is not driven by the presence of AGN in the sample we calculate the offset for 29 galaxies which show no signs of AGN activity, identifying a consistent median offset of 0.11\,$\pm$\,0.03\,dex\footnote{\tcr{We note if we adopt the more conservative \NH\,$\gtrsim$\,0.8 AGN criteria resulting in a 52\% AGN fraction (Sec. \ref{Sec:AGN}), we establish a slightly smaller offset of 0.08\,$\pm$\,0.03 dex.}}.}

Again, focussing on the brightest (log$_{10}$(L$_{\rm FIR}$[L$_{\odot}$])\,$>$12.0) and faintest (log$_{10}$(L$_{\rm FIR}$[L$_{\odot}$])\,$<$11.5) far-infrared \FIRS\ galaxies, we establish a median offset of 0.15\,$\pm$\,0.03\,dex and  0.08\,$\pm$\,0.04\,dex respectively. This indicates that at fixed stellar mass, for the high star-formation rates of the most far-infrared luminous \FIRS\ galaxies, the FMR predicts lower gas-phase metallicities than observed. This cautions against using the mass metallicity or fundamental metallicity relations to infer the metallicity, as is often routine in studies of luminous far-infrared galaxies, where direct metallicity measurements are unavailable \citep[e.g.][]{Magdis2012,Liu2019,Donevski2020,Gomez2022}

The elevated metallicity could be explained by a lack inflow of pristine gas into the galaxies, that would normally dilute the metal content of the interstellar medium, although this would also result in a short-lived star-formation episode, with no renewed fuel for star formation. An alternative scenario is that the metal enrichment and dust production in most far-infrared luminous \FIRS\ galaxies, is on average, higher than typical main-sequence galaxies, but it is unlikely to be the galaxies have higher gas fractions, given this would place them back on the fundamental metallicity relation. 

To explore these two scenarios, we now investigate the gas fraction of the \FIRS\ galaxies. Using the gas-phase metallicity and the dust mass we can infer the gas fraction (M$_{\rm g}$/M$_{\ast}$) of the \FIRS\ galaxies from the relations between dust to gas ratio (DtG) and metallicity. We use the non-redshift evolving, metallicity dependent, dust to gas ratio from \citet{Popping2022}, given by
\begin{equation}\label{Eqn:GDR}
   \rm  \log_{10}(DtG)\,=\,1.30\,\times\,(12+\log_{10}[O/H])-13.72 
\end{equation}
For the \FIRS\ sample, we measure a median dust to gas ratio of log$_{10}$(DtG)\,=\,$-$2.37\,$\pm$\,0.03 with a 16\qth\,--\,84\qth percentile range of log$_{10}$(DtG)\,=\,$-$2.38 to $-$\,2.34. In Figure \ref{fig:dFMR_MG}b, we show the gas fraction of the \FIRS\ galaxies as function of redshift, coloured by far-infrared luminosity. We include other literature samples of SFGs from the IRAM Plateau de Bure high-z blue sequence CO 3-2 survey  \citep[PHIBBS;][]{Tacconi2013,Tacconi2018}, the ASPECS survey \citep{Walter2016}, as well as CO-selected SMGs from \citet{Birkin2021} and, \Ha-detected SMGs from KAOSS (\citealp{Birkin2022}, Taylor et al. in prep.), for which we use Equation \ref{Eqn:GDR} to derive the gas fraction. On average, the  \FIRS\ galaxies have higher gas fractions than the \tcr{CO line luminosity derived} relation derived from \citet{Birkin2021}, with a median gas fraction of $\rm log_{10}(M_{g}/M_{\ast})$\,=\,0.26\,$\pm$\,0.08 with a 16\qth\,--\,84\qth percentile range of  $\rm log_{10}(M_{g}/M_{\ast})$\,=\,$-$0.23 to 0.62. \tcr{likely due to the inhomogeneous far-infrared selection of the \FIRS\ sample in contrast to the uniform 870\um\ selected \citet{Birkin2021} sample and systematic differences in methodology used to derived gas-masses.} \tcr{The \FIRS\ sample is} more consistent with the observationally derived main sequence scaling from \citet{Kokorev2021} (green lines in Figure \ref{fig:dFMR_MG}b), although we note this is derived using a different metallicity and gas to dust ratio calibration. This implies the median depletion timescale (t$_{\rm depl}$\,=\,M$_{\rm g}$/SFR), for the \FIRS\ sample is t$_{\rm depl}$[Gyr]\,=\,0.77\,$\pm$\,0.08, which is higher than the  t$_{\rm depl}$[Gyr]\,=\,0.21\,$\pm\,$0.04  reported by \citet{Birkin2021} for CO-selected SMGs. We note, in part this is likely driven by the broad range of far-infrared luminosity in the \FIRS\ galaxies, in addition to the lower, metallicity inferred, dust to gas ratio of our sample, compared to the typically adopted Milky Way value of log$_{10}$(DtG)\,=\,$-$2.0 \citep[e.g.][]{Swinbank2014,Birkin2022}

An alternative approach to deriving the gas mass of high-redshift galaxies, is to measure their CO line luminosity, and convert this to a gas mass using the  CO-to-H$_2$ conversion factor ($\alpha_{\rm CO}$) \cite[e.g.][]{Greve2005,Bothwell2013,Tacconi2018,Birkin2021,Rosenthal2026}. The choice of $\alpha_{\rm CO}$ has a significant impact on the derived gas mass and is a strong function of metallicity, with values varying from $\alpha_{\rm CO}$\,=\,1, consistent with local merging (U/LIRGs) galaxies, to $\alpha_{\rm CO}$\,=\,3.6, consistent with local disk galaxies (see \citealt{Boletta2013} for a full review). Whilst we do not have measurements of the CO line luminosity for the \FIRS\ sample, we can use the gas-phase metallicity to infer $\alpha_{\rm CO}$. We adopt the relation from \citet{Magdis2012}, that relates the gas-phase metallicity to $\alpha_{\rm CO}$ of the form,
\begin{equation} \label{Eqn:aCO}
       \rm  \log_{10}(\alpha_{\rm CO})\,=\,\alpha\,\times\,(12+\log_{10}[O/H]_{PP04})+\beta
\end{equation}
where the gas-phase metallicity is derived using the \citet{Pettini2004} \NH\ strong-line calibration, $\alpha$\,=\,$-$1.39\,$\pm$\,0.3 and $\beta$\,=\,12.8\,$\pm$\,2.2. For the \FIRS\ sample we derive a median 12+$\log_{10}$[O/H]$_{\rm PP04}$\,=\,8.76\,$\pm$\,0.02, which is slightly higher than that derived using Equation \ref{Eq:Bian}. Using Equation \ref{Eqn:aCO} we thus derive a median $\alpha_{\rm CO}$\,=\,4.12\,$\pm$\,0.31 with a 16\qth\,--\,84\qth\ percentile range of $\alpha_{\rm CO}$\,=\,2.73\,--\,6.31. We further identify no significant difference between far-infrared bright and faint galaxies with median values CO-to-H$_2$ conversion factor of $\alpha_{\rm CO}$\,=\,3.74\,$\pm$\,0.40 and $\alpha_{\rm CO}$\,=\,4.73\,$\pm$\,1.18. This implies that the CO-to-H$_2$ conversion factor of our sample is more consistent with local disk galaxies \citep[e.g.][]{leroy2011,Chiang2024}, and does not have strong dependence on far-infrared luminosity, or in fact main sequence offset. Adopting an  $\alpha_{\rm CO}$\,$\simeq$\,4 for the \citet{Birkin2021} sample of CO-selected SMGs, would move them to higher gas fractions, more aligned with the \FIRS\ sample \tcr{(see Figure \ref{fig:dFMR_MG}b)}. We note however that this inference of $\alpha_{\rm CO}$ is purely based on metallicity, which  has been demonstrated to likely be an over simplification, with  $\alpha_{\rm CO}$ having further dependences on the density and pressure of the interstellar medium \citep[e.g.][]{Desika2012,Accurso2017,Madden2020,Bisbas2025}.

\begin{figure*}[!htbp]
    \centering
    \includegraphics[width=\linewidth]{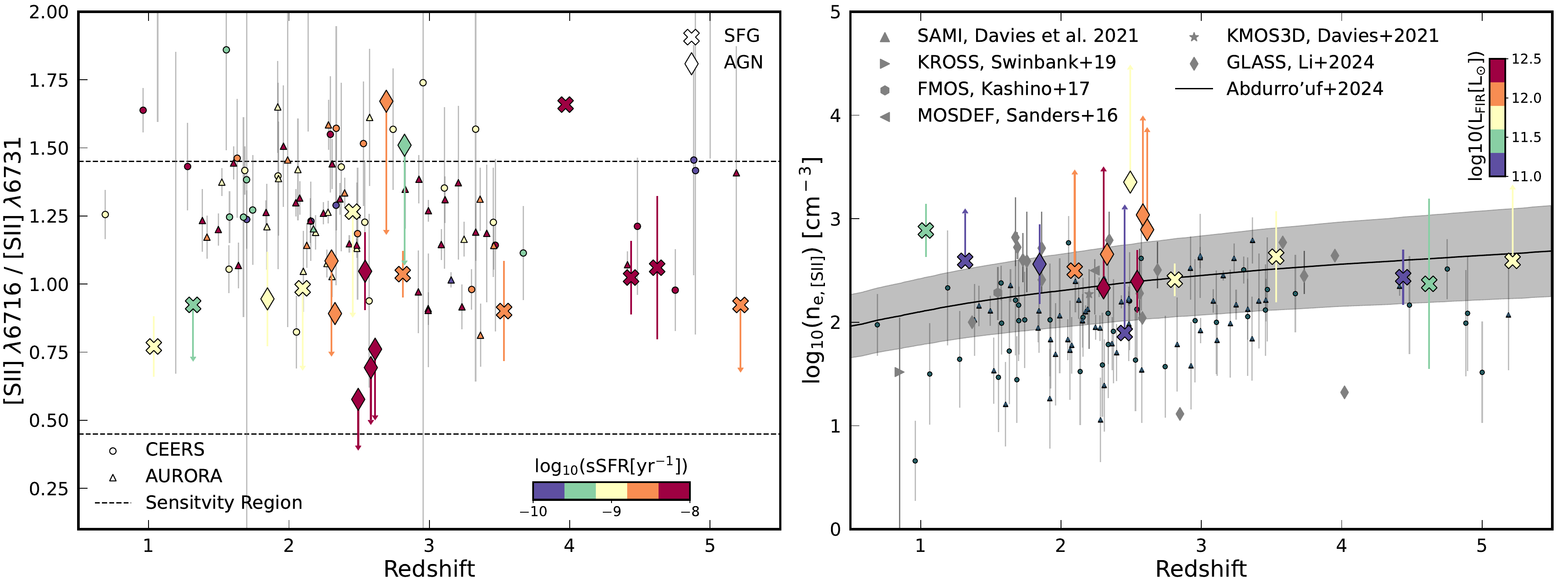}
    \caption{{Left:} \sii\ doublet emission-line ratio as function of spectroscopic redshift for the \FIRS\ sample coloured by specific star-formation rate (sSFR) \tcr{with diamonds indicating AGN and crosses SFGs}. Sources with S/N$<$3 in either \sii\ emission-lines are shown as limits.
    We show comparison samples of typical star-forming galaxies from the CEERS Survey \citet{Shapley2023} \tcr{(small circles)} and AURORA survey \citet{Shapley2025} \tcr{(small triangles)}. We further indicate the sensitivity region (black dashed lines) of the \sii\ line ratio to the electron density \citep{Kewley2019}. The \FIRS\ sample exhibits slightly lower \sii\ doublet emission-line ratios to the lower mass SFGs samples from CEERS and AURORA, with no clear dependence on sSFR.
    {Right:} The \sii\ derived electron density, using the \texttt{mappings v5.1} photoionisation code \citep{Kewley2019}, as a function of spectroscopic redshift, with the \FIRS\ galaxies coloured by dust mass \tcr{with diamonds indicating AGN and crosses SFGs}. The grey solid line (\tcr{and shaded region}) indicates the typical evolution (\tcr{and uncertainty}) of star-forming galaxies ($n_e$\,$\propto$\,(1+z)$^{1.2}$) from \citet{Abd2024}. The electron density of the \FIRS\ galaxies are consistent with typical star-forming galaxies, with no strong correlations with specific star-formation rate or far-infrared luminosity.}
    \label{fig:ne}
\end{figure*}

\subsection{Electron density}
In addition to the gas-phase metallicity and gas fractions, we can constrain the electron density within interstellar medium of the \FIRS\ galaxies,  using the \sii\, doublet ratio, allowing a physically motivated interpretation of the ionised interstellar medium conditions.  For the sample of \FIRS\ sample, we obtain detections (S/N$>$3) of the [SII] $\lambda$6716 emission in 17\% tcr{(8/48) and detections of the [SII] $\lambda$6731 emission line in 31\% (15/48) sources. For those sources with S/N$<$3 in either \sii\ line, we define a 3$\sigma$ upper limit, whilst for 30 sources with S/N$<$3 in both \sii\ lines, we do not derive the doublet ratio and electron density. We derive a median doublet ratio of \sii\,$\lambda$6716/\sii\,$\lambda$6731\,=\,0.98\,$\pm$\,0.05 and 16\,--\,84th percentile range of  \sii\,$\lambda$6716\,/\,\sii\,$\lambda$6731\,=\,0.77\,--\,1.30.  

In Figure \ref{fig:ne}a we plot the observed \sii\ doublet ratio as a function of spectroscopic redshift for the \FIRS\ galaxies as well as the CEERS and AURORA redshift matched samples, highlighting the region in which the ratio is sensitive to the electron density. The \FIRS\ galaxies exhibit slightly lower \sii\ doublet ratios to the SFGs from CEERS and AURORA at the same epoch, which have a median value of \sii\,$\lambda$6716/\sii\,$\lambda$6731\,=\,1.26\,$\pm$\,0.02. Previous studies of high-redshift SFGs have investigated the correlation between electron density and specific star-formation rate, often reporting mixed results \citep[e.g.][]{Kashino2019,Davies2021,Gillman2022,Topping2025}. In Figure \ref{fig:ne}a, we colour both the \FIRS\ galaxies and comparison samples by total specific star-formation rate (sSFR), where SFR$_{\rm tot}$ is defined as \Ha\ dust uncorrected SFR plus SFR$_{\rm FIR}$. We identify no clear correlation between \sii\ doublet ratio and sSFR.

Following the prescription of \citet{Kewley2019}, based on the \texttt{mappings v5.1}\footnote{\url{https://mappings.anu.edu.au/}} photoionisation models, we can convert the observed \sii\ doublet ratio to an electron density. For the \FIRS\ galaxies we derive a median electron density of $\log_{10}(n_{\rm e})$[cm$^{-3}$]\,=\,2.53\,$\pm$\,0.07 with a 16\,--\,84th percentile range of $\log_{10}(n_{\rm e})$[cm$^{-3}$]\,=\,2.05\,--\,2.81.
In Figure \ref{fig:ne}b, we show the electron density of the \FIRS\ galaxies, derived from the \sii\ doublet ratio as a function of spectroscopic redshift, compared to the CEERS and AURORA samples as well as literature studies of SFGs across a broad range of redshift from SAMI \citep[$z$\,$\sim$\,0;][]{Davies2021}, KROSS \citep[$z$\,$\sim$\,1;][]{Swinbank2019}, FMOS \citep[$z$\,$\sim$\,1.6;][]{Kashnino2017}, MOSDEF \citep[$z$\,$\sim$\,2.3;][]{Sanders2016} and, KMOS-3D \citep[$z$\,$\sim$\,2.6;][]{Davies2021} surveys. We include the evolution of electron density with redshift from \citet{Abd2024} of $n_e$\,$\propto$\,(1+z)$^{1.2}$, which predicts a median value of $\log_{10}(n_{\rm e})$[cm$^{-3}$]\,=\,2.41\,$\pm$\,0.03 and 16\,--\,84th percentile range of $\log_{10}(n_{\rm e})$[cm$^{-3}$]\,=\,2.01\,--\,2.79, across the redshift range of the \FIRS\ sample ($z$\,=\,1.04\,--\,5.22), consistent with the measured value. We note the large fraction of lower limits in Figure \ref{fig:ne}b which consistently fall above the relation from \citet{Abd2024}, potentially indicate higher signal to noise spectra could reveal elevated electron densities.

We further find no strong correlation with far-infrared luminosity in the \FIRS\ galaxies, indicating consistent electron densities independent of far-infrared emission, consistent with recent results from \citet{Kiyota2026}. We note that the electron density, and those of the comparison samples, are purely derived from rest-frame optical emission lines, which is not equivalent to that derived from far-infrared fine structure lines \citep[e.g.][]{Lee2019,Zhang2019} which tend to find lower electron densities ($\log_{10}(n_{\rm e})$[cm$^{-3}$]\,$<$\,2). This is attributed to the fine structure lines being sensitive to different densities and physical conditions within the interstellar medium, than the optical diagnostics \citep[e.g.][]{Schimek2024,Harikane2025,Usui2025}. 

\subsection{Ionisation parameter}

\begin{figure}[!htbp]
    \centering
    \includegraphics[width=\linewidth]{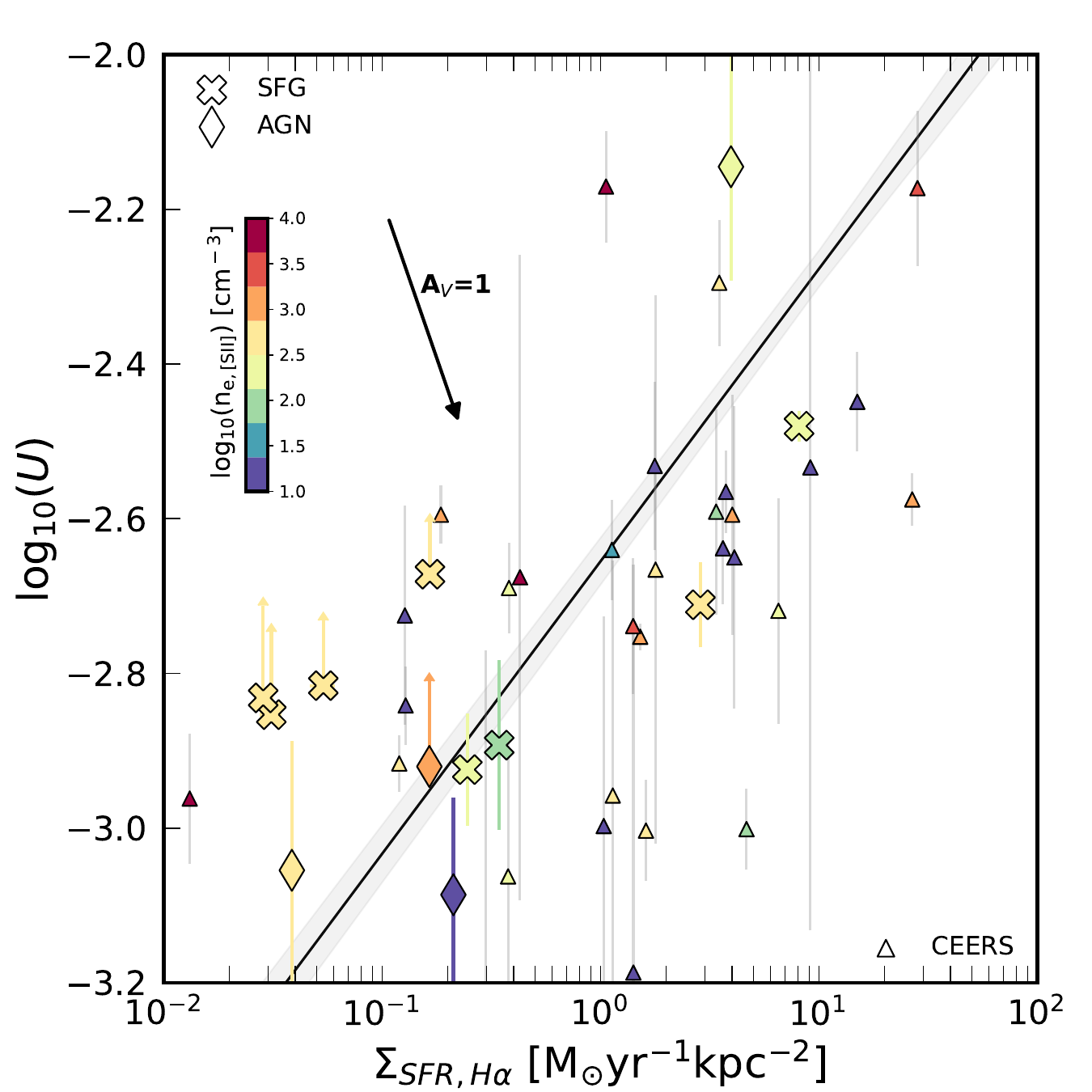}
    \caption{Ionisation parameter ($U$), as derived from the O$_{32}$ strong-line ratio, as a function of \Ha\ star-formation rate surface density \tcr{with diamonds indicating AGN and crosses SFGs from the \FIRS\ sample}. For the \FIRS\ galaxies this represents the observed (dust-uncorrected) values whilst for the CEERS sample \tcr{(small triangles)} we show the intrinsic (dust-corrected) values. We colour the galaxies by \sii\ derived electron density and show the parametric fit from \citet{Reddy2023} for the CEERS survey of the form log$_{10}(U)$\,=\,(0.378\,$\pm$\,0.021)$\times$log$_{10}$(sSFR)$-$(2.655$\pm$0.011). The black arrow indicates the impact of dust attenuation with A$_V$\,=\,1. The observed (dust-uncorrected) values of the \FIRS\ galaxies are consistent with dust-corrected CEERS galaxies, although with low number statistics.}
    \label{fig:USFR}
\end{figure}

We can further constrain the ionisation conditions of the interstellar medium of the \FIRS\ galaxies through the ionisation parameter ($U$), which represents the ratio of the density of ionising photons to the density of hydrogen atoms \citep{Osterbrock2006}. Of the rest-frame optical emission lines detected in the NIRSpec spectra, the O$_{32}$ strong-line ratio (\oiii\,5007$\lambda$\,/\oii\,3727,9$\lambda$$\lambda$) is sensitive to the ionisation parameter, where higher ratios O$_{32}$, at fixed metallicity and interstellar medium pressure, indicate a more highly ionised nebula.

Utilising the numerical prescription from the \texttt{mappings v5.1} photoionisation code \citep{Kewley2019b}, and the gas-phase metallicity for each source as given by Equation \ref{Eq:Bian}, at a constant interstellar medium pressure of log$_{10}$(P/k)\,=\,5\footnote{We note a higher pressure would result in higher ionisation parameter for a given metallicity, see \citet{Kewley2019b} for details.} we can derive the ionisation parameter for the \FIRS\ galaxies. In 25\% (12/48) of the sample we have coverage of the \oii\,3727,9$\lambda\lambda$ emission line, with five sources having S/N$<$3, thus we define a 3$\sigma$ upper limit. Given the wavelength separation of the \oii\,3727,9$\lambda\lambda$ and \oiii\,5007$\lambda$ emission lines, to derive an accurate O$_{32}$ index, and thus ionisation parameter, it is critical to dust correct both emission lines. However, given the lack of coverage of the \Hb\ emission-line in the \FIRS\ galaxies, and thus poorly constrained nebular attenuation, we first derive the observed (dust-uncorrected) ionisation parameter for the sources. This represents an upper limit on the intrinsic ionisation parameter of the sources, given a nebula dust correction would act to reduce the O$_{32}$ index, and thus reduce the ionisation parameter further.

We derive a median (dust-uncorrected) O$_{32}$ index of O$_{32}$\,=\,0.70\,$\pm$\,0.10 with a 16\qth\,--\,84\qth of  O$_{32}$\,=\,0.54\,--\,1.13 which corresponds to a median (dust-uncorrected) ionisation parameter of $\log_{10}(U)$\,=\,$-$2.84\,$\pm$\,0.06 with a 16\,--\,84th percentile range of  $\log_{10}(U)$\,=\,$-$2.96 to $-$2.63. This is much lower than expected given the high star formation rates of DSFGs \citep[e.g.][]{Casey2014,Dud2020} and is comparable to that derived for the typical star-forming galaxies at the same epoch \citep[e.g.][]{Sanders2016,Kaasinen2018,Reddy2023}.

It is well known for SFGs, that there is strong correlation between ionisation parameter and \Ha\ star-formation rate surface density ($\Sigma_{\rm H\alpha, SFR}$), whereby galaxies with higher star formation rate surface densities have an increased ionisation parameter i.e. a greater amount of ionising photons to the density of hydrogen atoms in the interstellar medium. \citep[e.g.][]{Nakajima2014,Kaasinen2018,Papovich2022,Reddy2023}. However, whether this relation holds true for dust-obscured and star-bursting systems is unknown. In Figure \ref{fig:USFR} we plot the observed (not dust corrected) ionisation parameter ($U$), as derived from the O$_{32}$ index as a function of \Ha\ star-formation rate surface density ($\Sigma_{\rm H\alpha, SFR}$) derived using the \FIRS\ galaxies optical sizes (see Sec. \ref{Sec:Sizes}), as well as the parametric scaling derived by \citet{Reddy2023} for the CEERS survey.

We further show the individual (dust-corrected) measurements from the redshift matched sample of CEERS galaxies, derived using the same methodology. For the CEERS galaxies we further utilise the observed \Hab\ ratio to dust correct the emission line fluxes, and thus derive a dust corrected ionisation parameter and \Ha\ star-formation rate surface density. Assuming Case B recombination at an electron temperature of $T_{\rm e}$\,=\,10,000\,K and an electron density of $n_{\rm e}$\,=\,100\,cm$^{-3}$ \citep{Osterbrock2006}\footnote{We note at fixed electron temperature, adjusting the electron density to observed value does not alter the intrinsic Balmer decrement.}, the nebular dust attenuation  derived from the Balmer decrement is defined as,
\begin{equation}
A_{V,\mathrm{gas}} = R_V E(B - V) = \frac{R_V}{0.4 \left( K_{\mathrm{H}\beta} - K_{\mathrm{H}\alpha} \right)} \times \log_{10} \left( \frac{\mathrm{H}\alpha / \mathrm{H}\beta}{2.86} \right)
\end{equation}
where $K_{\lambda}$ denotes the magnitude of extinction at a given wavelength from the extinction curve, and $R_{\rm V}$ is the total-to-selective extinction ratio. For the CEERS sample the median Balmer decrement is \Hab\,=\,3.22\,$\pm$\,0.16 and thus with $R_{\rm V}$\,=\,4.05 the median \av\,=\,0.6\,$\pm$\,0.1 

For the \FIRS\ galaxies, given the inconsistent coverage of the \Hb\ emission line we can not derive the dust correction for all sources. The \FIRS\ galaxies scatter about the relation indicating a similar trend maybe present in DSFGs. Previous studies of star-forming galaxies have further identified that galaxies with high ionisation parameters exhibit systematically lower electron densities \citep[e.g.][]{Kashino2019,Reddy2023b,Topping2025}. In Figure \ref{fig:USFR} we colour both the CEERS sample and \FIRS\ galaxies by their \sii\ derived electron density, although given the low number statistics, it is not clear whether the \FIRS\ galaxies follow this trend. 

As indicated by the black arrow in Figure \ref{fig:USFR}, even with an A$_{V,\,{\rm Nebular}}$\,=\,1, there is a strong correction on both the ionisation parameter and \Ha\ star-formation rate surface density when the dust content is taken into account. For nine sources, we have detections of both the O$_{32}$ ratio and the \Hb\ line with a median Balmer decrement of \Hab\,=\,5.84\,$\pm$\,0.75, thus adopting a \citet{Calzetti2000} attenuation curve with $R_{\rm V}$\,=\,4.05 and electron density of $n_{\rm e}$\,=\,100\,cm$^{-3}$ (consistent with Figure \ref{fig:ne}), we derive a nebular dust attenuation of A$_{V,\,{\rm Nebular}}$\,=\,2.55\,$\pm$\,0.43\footnote{We note this is an upper limit on the nebular attenuation as we do not account for potential stellar absorption from older stellar populations.}. Whilst this is for only a subset of sources we expect A$_{V,\,{\rm Nebula}}$\,$>$\,1 in the majority \FIRS\ galaxies, given the expectation of higher nebular attenuation than stellar attenuation, especially in massive galaxies \citep[e.g.][]{Calzetti2000,Reddy2020,Lorenz2026} and 
from Table \ref{Table:SED}, the median \texttt{bagpipes} derived stellar attenuation of the sample is A$_{V,\,{\rm Stellar }}$\,$=$\,2.23\,$\pm$\,0.14.  

For the nine sources with detections of both the O$_{32}$ ratio and \Hb\ line we establish a median dust-corrected ionisation parameter of  $\log_{10}(U)$\,=\,$-$3.93\,$\pm$\,0.37. However, we note this is an upper limit given the photoionization models presented in \citet{Kewley2019b} are not defined below $\log_{10}(U)$\,$<$\,$-$3.9. This would result in the \FIRS\ galaxies lying below the relation of typical SFGs in Figure \ref{fig:USFR}, with lower ionisation parameters than implied by their high star-formation rate surface densities \citep[e.g.][]{Kashino2019}. The dust corrected ionisation parameter is derived assuming the \FIRS\ galaxies have a similar dust attenuation law to \citet{Calzetti2000} which many studies have suggested otherwise \cite[e.g.][]{Kriek2013,Popping2017,Hamed2023,Sun2026}. Furthermore, the optical line O$_{32}$ ratios dependence on the ionisation parameter is a strong function of interstellar medium pressure and only traces the less obscured, ionised, regions of the galaxies, in contrast to the far-infrared fine structure lines which are sensitive to deeply obscured star formation \citep[e.g.][]{Rig2018,Decarli2025}. The optical line diagnostics further have potential contamination from diffuse ionised gas which local studies have shown can have significantly lower the ionisation parameter \citep[e.g.][]{Zhang2017, Belifore2022}. 


\section{Conclusions}\label{Sec:Conc}

We present a spectroscopic analysis of a comprehensive, but inhomogeneously selected, sample identified from archival ALMA observations \citep{Liu2019,A3GOODS2024,Lee2025}. Utilising the NIRSpec median resolution (or higher) observations from DJA, from an initial sample of 121 galaxies, we built a final sample of 48 \FIRS\ galaxies with robust ($>$3$\sigma$) ALMA counterparts and rest-frame optical spectroscopy that captures the \Ha\ emission-line. We determine the samples multi-wavelength (near-infrared and far-infrared) SED properties as well as defining the galaxies interstellar medium properties from rest-frame optical diagnostics. Our main conclusions are:

\begin{itemize}
     \item  Star formation and dust content: On average the \FIRS\ sample lies 0.32\,$\pm$\,0.07\,dex above the main sequence with the highest infrared luminosity galaxies lying at $\gtrsim$0.5\,dex, consistent with other far-infrared bright galaxies from ASUDS \citep{Dud2020} and KAOSS \citep{Birkin2023}, as well as the most massive, sub-millimetre bright galaxies, from L-Galaxies \citep{Araya2025a}. 
     The \FIRS\ galaxies have median dust fraction of log$_{10}$(M$_{\rm d}$/M$_{\rm \ast}$)\,=\,$-$2.20\,$\pm$\,0.08, which as shown in Figure \ref{fig:dMS_Md}, is consistent with other the sub-millimetre bright galaxy samples at a median redshift of $z$\,=\,2.5, but the sample exhibits higher dust fractions at higher redshift.  
     
     \item Ionisation origin: The majority of the sample (29/48) show no signs of AGN activity in terms of X-ray counterparts, optical emission-line ratios (Figure \ref{fig:BPT}) with an average \NH\ index of \NH\,=\,0.61\,$\pm$\,0.06, or broad emission line features with an average \Ha\ FWHM of 328\,$\pm$\,27 km/s. Of the 19 galaxies flagged as AGN, 15 sources have X-ray counterparts $>$10$^{42}$\,erg/s, 3 occupy the AGN region of the BPT diagram at their epoch, based on the \citet{Kewley2013} redshift evolving demarcation, and nine galaxies show improved emission-line fits ($\Delta$BIC$_{\rm H\alpha,[\textsc{nii}],[\textsc{sii}]}$\,$<$$-$10) when a broad component is included, with a median \Ha\ broad FWHM of 1818\,$\pm$\,241 km/s. Although, we note this is a lower limit on the AGN activity in the sample given the stochastic placement of NIRSpec/MSA slits that can miss the central, core, regions of galaxies, which may harbour AGN.
     
    \item Metallicity and gas fraction: We identify super solar, gas-phase metallicities in \FIRS\ galaxies compared to typical SFGs at the same epoch and other \Ha-detected high-redshift SMG samples, with a median gas-phase metallicity of \ZOH\,=\,8.71\,$\pm$\,0.02 derived from the \NH\, index using the \citet{Bian2018} calibration. For far-infrared bright (log$_{10}$(L$_{\rm IR}$[L$_{\odot}$])\,$>$\,12.0) galaxies we derive a median metallicity offset to the \citet{Curti2020} fundamental metallicity plane of 0.15\,$\pm$\,0.03\,dex, suggesting DSFGs do not follow the same equilibrium between metal enrichment and star formation as more typical, less dust-obscured, galaxies.
    Using the metallicity dependent gas-to-dust ratio from \citet{Popping2017},  we derive a median gas fraction of  $\rm log_{10}(M_{g}/M_{\ast})$\,=\,0.26\,$\pm$\,0.08, which is elevated with respect to other studies of bright SMGs at the same epoch,   suggesting a strong connection between far-infrared luminosity, metal content and gas fraction.
    
    \item {Electron density}: From an analysis of the \sii\ emission-line doublet ratio, and adopting the photoionization modelling from \citet{Kewley2019}, we established a median electron density for the \FIRS\ sample of $\log_{10}(n_{\rm e})$[cm$^{-3}$]\,=\,2.53\,$\pm$\,0.07, consistent with star-forming galaxies at the same epoch and the cosmic evolution of electron density as parametrised by \citet{Abd2024}, with no strong correlation with specific star-formation rate or far-infrared luminosity.
    
   \item {Ionisation parameter}: Utilising the sensitivity of the O$_{32}$ index to the ionisation parameter ($U$), and following the prescriptions of \citet{Kewley2019b}, we derive a median (dust-uncorrected) ionisation parameter of $\log_{10}(U)$\,=\,$-$2.84\,$\pm$\,0.06 for a subsample of the \FIRS\ galaxies, comparable to the dust-corrected ionisation parameters of the typical star-forming galaxies at the same epoch. Finally, we derive an upper limit on the dust-corrected ionisation parameter ($\log_{10}(U)$\,=\,$-$3.93\,$\pm$\,0.37) suggesting the far-infrared detected galaxies fall below the ionisation parameter to \Ha\ derived star-formation rate relation established for star-forming galaxies when using optical line diagnostics and a \citet{Calzetti2000} dust attenuation law.

\end{itemize}

Despite the dust rich nature of the \FIRS\ sample, our analysis indicates that the electron density, as derived from optical line diagnostics, are consistent with the broader population of star-forming galaxies at the same epoch. However, the inferred ionisation parameters are lower than expected given the rate of star formation rates of the sample, whilst the brightest far-infrared sources exhibit elevated gas-phase metallicities relative to their stellar masses and star-formation rates. These findings suggest the \FIRS\ sources do not maintain the same equilibrium between metal enrichment, dust production and star formation as more typical, less dust-obscured, galaxies. Caution is therefore needed when adopting the mass metallicity to infer the metallicity of DSFGs. 

While our study demonstrates the power of NIRSpec/MSA spectroscopy in detecting the rest-frame optical emission line features in the previously inaccessible parameter space of high-redshift DSFGs, limitations remain. Non-optimum placement of NIRSpec slits and heterogenous signal to noise across the archival observations introduces potential biases that affect our derived properties. Future higher signal to noise NIRSpec observations will be essential to robustly characterise the multi-phase interstellar medium of dust-obscured galaxies.

\section{Data availability} \label{App:FIR}
A machine readable table of the 138 sources with spectroscopic redshifts and ALMA detections, in addition to the \texttt{mercurius} MCMC fitting results, and a visualisation of the NIRSpec spectra, for the final sample can be found on this \href{https://doi.org/10.5281/zenodo.18712702}{link}}. Tables \ref{App:Selection}, \ref{App:NIRSpec_table} \& \ref{App: SED_table}, are only available in electronic form at the CDS via anonymous ftp to cdsarc.u-strasbg.fr (130.79.128.5) or via http://cdsweb.u-strasbg.fr/cgi-bin/qcat?J/A+A/.

\begin{acknowledgements}
The observations analysed in this work are made with the NASA/ESA/CSA James
Webb Space Telescope (DOI: 10.17909/z7p0-8481). We thank Ryan Sanders and Alice Shapley for sharing the NIRSpec properties for the CEERS and AURORA surveys.
We acknowledge financial support from the Cosmic Dawn Center (DAWN), funded by the Danish National Research Foundation (DNRF) under grant  DNRF140.
FV, KI, AP acknowledges support from the Independent Research Fund Denmark (DFF) under grant 3120-00043B.
BG, CH and UD acknowledge support from Villum Fonden under project grant 53120.
PA-A acknowledge support from the Independent Research Fund Denmark (DFF) under grant 4251-00086B.
SJ acknowledges the Villum Fonden research grants 37440 and 13160.
PFW acknowledges funding through the National Science and Technology Council grants 113-2112-M-002-027-MY2.
DC is supported by research grant PID2024-156100NB-C21  financed by MICIU/AEI /10.13039/501100011033 / FEDER, EU., and the research grant CNS2024-154550 funded by MI-CIU/AEI/10.13039/501100011033.
TK acknowledges support from JSPS grant 25KJ1331.
FR acknowledges support from the Dutch Research Council (NWO) through the Veni grant VI.Veni.222.146. 
SF acknowledges support from the Dunlap Institute, funded through an endowment established by the David Dunlap family and the University of Toronto.
JRW acknowledges that support for this work was provided by The Brinson Foundation through a Brinson Prize Fellowship grant.
MO acknowledges support from JSPS KAKENHI Grant Number JP25K07361. ML acknowledges support from the European Union’s Horizon 2020 research and innovation program under the Marie Skłodowska-Curie grant agreement No 101107795 and by the DeiC National HPC (DeiC-DTU-N2-2024057).

\end{acknowledgements}

\bibliography{master}{}
\bibliographystyle{aa}

\begin{appendix}
\FloatBarrier
\begin{table*}
\section{Far-infrared selection}\label{App:Selection}
\centering
\renewcommand{\arraystretch}{1}
\begin{tabular}{ccccccccc}
\hline
NIRSpec ID & ALMA ID & R.A. & Dec. & $z_{\rm spec}$ & Flux & Band & Sep\\
& &[deg.] & [deg.] & & [mJy] &  & ["]\\
\hline
RB-160718 & A2-424.0 & 34.480165 & $-$5.112523 & 3.5444 & 6.10$\pm$0.70 & 7 & 0.13 \\
GT-167859 & A3C-941026 & 150.163540 & 2.372423 & 2.0836 & 7.20$\pm$0.36 & 7 & 0.35 \\
$^{\dagger\ddagger*}$GR-172113 & A3G-9834 & 53.148856 & $-$27.821183 & 2.5829 & 4.22$\pm$0.12 & 7 & 0.01 \\
$^{\ddagger}$JD-159764 & A3G-15876 & 53.160581 & $-$27.776188 & 2.5428 & 0.55$\pm$0.09 & 6 & 0.19 \\
$^{\dagger}$DD-160970 & A2-38.0 & 34.778015 & $-$5.366046 & 2.5448 & 7.70$\pm$0.30 & 7 & 0.09 \\
$^{\dagger}$BA-173854 & A3C-892054 & 150.098507 & 2.365359 & 2.5554 & 0.42$\pm$0.12 & 4 & 0.09 \\
$^{\ddagger}$JD-172437 & A3G-16972 & 53.081969 & $-$27.767276 & 2.3032 & 1.59$\pm$0.23 & 6 & 0.17 \\
DD-147398 & A3C-872955 & 149.908646 & 2.115584 & 2.9436 & 4.84$\pm$0.71 & 7 & 0.07 \\
GR-172160 & A3G-15639 & 53.181371 & $-$27.777579 & 2.6974 & 4.94$\pm$0.84 & 8 & 0.02 \\
GT-162506 & A2-125.0 & 34.363199 & $-$5.199367 & 2.1562 & 4.60$\pm$0.50 & 7 & 0.35 \\
GT-162118 & A2-272.0 & 34.461197 & $-$5.201328 & 1.7059 & 5.10$\pm$0.50 & 7 & 0.05 \\
JD-173408 & A3G-3818 & 53.074866 & $-$27.875907 & 4.4319 & 0.09$\pm$0.02 & 4 & 0.06 \\
$^{\dagger\ddagger*}$GR-172166 & A3G-16274 & 53.131145 & $-$27.773187 & 2.2245 & 0.04$\pm$0.01 & 4 & 0.02 \\
$^{\dagger\ddagger*}$JD-149659 & A3G-10497 & 53.182003 & $-$27.814182 & 2.6138 & 1.03$\pm$0.29 & 7 & 0.05 \\
RB-151060 & A2-346.0 & 34.273508 & $-$5.251360 & 4.5440 & 3.60$\pm$0.30 & 7 & 0.13 \\
JD-170092 & A3G-54608 & 53.108837 & $-$27.869126 & 3.7476 & 1.45$\pm$0.19 & 6 & 0.28 \\
$^{\ddagger}$JD-167963 & A3G-6131 & 53.091556 & $-$27.853365 & 2.0346 & 2.76$\pm$0.50 & 7 & 0.29 \\
RB-161829 & A2-250.0 & 34.202863 & $-$5.121150 & 3.9706 & 3.30$\pm$0.60 & 7 & 0.06 \\
$^{\dagger\ddagger}$JD-176494 & A3G-15432 & 53.146178 & $-$27.779934 & 2.5834 & 0.23$\pm$0.06 & 6 & 0.03 \\
GT-169541 & A3C-893896 & 150.130915 & 2.208874 & 2.3032 & 0.11$\pm$0.03 & 4 & 0.27 \\
$^{\ddagger}$JD-164819 & A3G- & 53.091784 & $-$27.712084 & 1.6153 & 2.86$\pm$0.54 & 7 & 0.35 \\
$^{\dagger}$GR-175918 & A3G-20173 & 53.154439 & $-$27.738678 & 2.3279 & 0.81$\pm$0.25 & 6 & 0.06 \\
RB-163266 & A2-427.0 & 34.301094 & $-$5.287985 & 4.8041 & 2.40$\pm$0.50 & 7 & 0.08 \\
$^{\dagger\ddagger}$GT-162424 & EG-127505 & 34.337315 & $-$5.143674 & 4.5560 & 2.37$\pm$0.29 & 7 & 0.03 \\
BA-155180 & A3C- & 150.112224 & 2.313952 & 2.0977 & 1.08$\pm$0.21 & 6 & 0.45 \\
GR-172171 & A3G-45150 & 53.137103 & $-$27.761408 & 1.8130 & 0.84$\pm$0.04 & 6 & 0.08 \\
RB-157952 & EG-138976 & 34.276132 & $-$5.163512 & 5.2172 & 0.24$\pm$0.05 & 4 & 0.30 \\
JD-175480 & A3G-22281 & 53.097333 & $-$27.720451 & 2.5651 & 1.82$\pm$0.22 & 7 & 0.12 \\
$^{\dagger}$BA-173819 & A3C-765691 & 150.086472 & 2.264299 & 2.4934 & 0.14$\pm$0.04 & 6 & 0.13 \\
DD-150418 & A2-659.0 & 34.445445 & $-$5.130508 & 1.8187 & 1.70$\pm$0.30 & 7 & 0.08 \\
$^{\ddagger}$JD-163721 & A3G-7551 & 53.131446 & $-$27.841383 & 1.6145 & 0.43$\pm$0.03 & 4 & 0.03 \\
GT-169474 & A3C-481073 & 150.166891 & 2.235805 & 1.8404 & 0.42$\pm$0.09 & 6 & 0.04 \\
BA-173884 & A3C-320337 & 150.087998 & 2.395073 & 3.5328 & 1.34$\pm$0.16 & 7 & 0.14 \\
DD-149715 & A3C-77714 & 150.077170 & 2.380370 & 2.8107 & 0.44$\pm$0.03 & 6 & 0.05 \\
GT-173952 & A2-2.0 & 34.465044 & $-$5.251940 & 2.3039 & 1.02$\pm$0.18 & 7 & 0.27 \\
$^{\ddagger}$GT-175747 & EG-133443 & 34.415788 & $-$5.171871 & 1.9168 & 1.04$\pm$0.15 & 7 & 0.04 \\
$^{\ddagger}$GR-172157 & A3G-15381 & 53.180521 & $-$27.779717 & 2.6935 & 0.36$\pm$0.10 & 6 & 0.03 \\
GT-162594 & EG-129470 & 34.471635 & $-$5.256968 & 2.5264 & 0.96$\pm$0.16 & 7 & 0.05 \\
GR-175915 & A3G-20740 & 53.171807 & $-$27.733608 & 2.5750 & 0.06$\pm$0.01 & 4 & 0.25 \\
EX-172449 & EG-82011 & 34.350624 & $-$5.149878 & 4.6229 & 0.77$\pm$0.08 & 7 & 0.06 \\
GR-146584 & A3G-15336 & 53.179110 & $-$27.780619 & 1.0372 & 0.17$\pm$0.04 & 6 & 0.02 \\
$^{\ddagger}$JD-172117 & A3G-10832 & 53.161428 & $-$27.811149 & 2.8233 & 0.39$\pm$0.12 & 6 & 0.06 \\
JD-176095 & A3G-22642 & 53.052276 & $-$27.718373 & 1.8070 & 3.12$\pm$0.85 & 9 & 0.11 \\
GR-172150 & A3G-14638 & 53.176590 & $-$27.785521 & 1.3172 & 0.19$\pm$0.06 & 6 & 0.06 \\
GR-172162 & A3G-15844 & 53.170916 & $-$27.775426 & 2.4542 & 0.15$\pm$0.03 & 6 & 0.03 \\
$^{\ddagger}$JD-146516 & A3G- & 53.155657 & $-$27.779365 & 1.8469 & 0.12$\pm$0.02 & 6 & 0.18 \\
GR-172152 & A3G-15016 & 53.173693 & $-$27.782134 & 1.9981 & 0.10$\pm$0.02 & 6 & 0.07 \\
JD-172382 & A3G-15927 & 53.095744 & $-$27.774815 & 4.4374 & 0.26$\pm$0.07 & 7 & 0.17 \\
\hline

\end{tabular}
\tablefoot{
The source properties for the 48 far-infrared detected NIRSpec sources, ranked in descending infrared luminosity, as per Figure \ref{fig:RGB}. For each source we list the NIRSpec ID (Column 1, C1), ALMA ID, either AS2UDS (A2-) or A3COSMOS (A3-) or ECOGAL (EG-) (C3), J2000 R.A (C5), J2000 DEC. (C6) \tcr{of the NIRSpec source position}, NIRSpec spectroscopic redshift (C7), ALMA primary beam corrected flux (C8), ALMA Band, ALMA\,--\,NIRSpec separation (C10). The NIRSpec ID is given as JWST program acronym (DD\,=\,DeepDive, RB\,=\,Rubies, JD\,=\,JADES, GR\,=\,George Rieke, GT\,=\,GTO, BA\,=\,BlueJay) followed by the DJA object-id (\texttt{objid}). DJA spectra file names are included in the online version of the table. Sources with no ALMA ID originate from the \texttt{blind} detection catalogue of the respective ALMA archival program.   Sources identified as AGN (see Sec. \ref{Sec:AGN}) are highlighted, where $\dagger$ indicates broad emission lines, $\ddagger$ indicates an X-ray counterpart and $*$ highlights elevated optical-line ratios.
}
\end{table*}


\begin{table*}
\section{Emission-line properties}\label{App:NIRSpec_table}
\centering
\label{Table:NIRspec}
\renewcommand{\arraystretch}{1}
\begin{tabular}{clcccccccc}
\hline
NIRSpec ID & $z_{\rm spec}$ & \underline{H$\alpha$} & A$_{\rm V, Balmer}$ & \underline{[OIII 5007$\lambda$]} &   \underline{[NII]} & 12+log(O/H) & \underline{[SII 6716$\lambda$]} & log$_{10}$($n_{\rm e}$) \\
 & & H$\beta$ & & H$\beta$  &   H$\alpha$ & &[SII 6731$\lambda$] & [cm$^{\rm -3}$] \\
\hline
RB-160718 & 3.5444 & - & - & - & 0.55$\pm$0.15 & 8.69$\pm$0.06 & - & - \\
GT-167859 & 2.0836 & - & - & - & 0.36$\pm$0.08 & 8.61$\pm$0.05 & - & - \\
$^{\dagger\ddagger*}$GR-172113 & 2.5829 & $>$2.73 & - & $>$40.95 & $<$1.36 & $<$8.89 & - & - \\
$^{\ddagger}$JD-159764 & 2.5428 & 3.85$\pm$0.26 & 1.11$\pm$0.23 & 1.43$\pm$0.10 & 0.27$\pm$0.02 & 8.54$\pm$0.02 & 1.05$\pm$0.14 & 2.40$\pm$0.30 \\
$^{\dagger}$DD-160970 & 2.5448 & $>$2.36 & - & - & 0.81$\pm$0.19 & 8.77$\pm$0.05 & - & - \\
$^{\dagger}$BA-173854 & 2.5554 & $>$1.78 & - & - & $<$1.31 & $<$8.92 & - & - \\
$^{\ddagger}$JD-172437 & 2.3032 & - & - & - & 0.75$\pm$0.11 & 8.76$\pm$0.03 & $<$1.09 & $>$2.33 \\
DD-147398 & 2.9436 & $>$0.93 & - & $>$1.71 & $<$0.54 & $<$8.69 & - & - \\
GR-172160 & 2.6974 & $>$0.43 & - & - & 0.59$\pm$0.16 & 8.71$\pm$0.06 & - & - \\
GT-162506 & 2.1562 & - & - & - & 0.82$\pm$0.12 & 8.78$\pm$0.03 & - & - \\
GT-162118 & 1.7059 & - & - & - & 0.84$\pm$0.31 & 8.78$\pm$0.08 & - & - \\
JD-173408 & 4.4319 & - & - & - & 0.56$\pm$0.08 & 8.70$\pm$0.03 & - & - \\
$^{\dagger\ddagger*}$GR-172166 & 2.2245 & 12.44$\pm$5.66 & 5.17$\pm$1.18 & 7.15$\pm$2.85 & 1.34$\pm$0.57 & 8.88$\pm$0.09 & - & - \\
$^{\dagger\ddagger*}$JD-149659 & 2.6138 & $>$3.54 & - & $>$0.78 & 0.86$\pm$0.15 & 8.79$\pm$0.04 & $<$0.76 & $>$2.89 \\
RB-151060 & 4.5440 & - & - & - & 0.72$\pm$0.19 & 8.75$\pm$0.06 & - & - \\
JD-170092 & 3.7476 & $>$2.72 & - & - & 0.60$\pm$0.15 & 8.71$\pm$0.05 & - & - \\
$^{\ddagger}$JD-167963 & 2.0346 & - & - & - & 0.52$\pm$0.10 & 8.68$\pm$0.04 & - & - \\
RB-161829 & 3.9706 & - & - & - & 1.01$\pm$0.20 & 8.82$\pm$0.04 & - & - \\
$^{\dagger\ddagger}$JD-176494 & 2.5834 & - & - & - & 1.25$\pm$0.20 & 8.87$\pm$0.03 & $<$0.69 & $>$3.04 \\
GT-169541 & 2.3032 & - & - & - & $<$0.71 & $<$8.75 & - & - \\
$^{\ddagger}$JD-164819 & 1.6153 & 3.92$\pm$1.73 & 1.19$\pm$0.97 & 0.82$\pm$0.50 & 0.37$\pm$0.06 & 8.61$\pm$0.04 & - & - \\
$^{\dagger}$GR-175918 & 2.3279 & - & - & - & 0.95$\pm$0.08 & 8.81$\pm$0.02 & 0.89$\pm$0.00 & 2.66$\pm$0.00 \\
RB-163266 & 4.8041 & - & - & - & 1.74$\pm$0.68 & 8.94$\pm$0.08 & - & - \\
$^{\dagger\ddagger}$GT-162424 & 4.5560 & - & - & - & $<$0.71 & $<$8.75 & - & - \\
BA-155180 & 2.0977 & 6.40$\pm$0.58 & 2.86$\pm$0.32 & 0.39$\pm$0.07 & 0.33$\pm$0.03 & 8.59$\pm$0.02 & $<$0.98 & $>$2.50 \\
GR-172171 & 1.8130 & $>$2.88 & - & - & 0.62$\pm$0.12 & 8.72$\pm$0.04 & - & - \\
RB-157952 & 5.2172 & $>$4.85 & - & $>$1.31 & 0.42$\pm$0.04 & 8.63$\pm$0.02 & $<$0.92 & $>$2.59 \\
JD-175480 & 2.5651 & $>$1.45 & - & - & 0.57$\pm$0.12 & 8.70$\pm$0.05 & - & - \\
$^{\dagger}$BA-173819 & 2.4934 & 6.00$\pm$0.59 & 2.66$\pm$0.34 & - & 0.18$\pm$0.04 & 8.46$\pm$0.05 & $<$0.58 & $>$3.35 \\
DD-150418 & 1.8187 & - & - & - & 0.69$\pm$0.11 & 8.74$\pm$0.03 & - & - \\
$^{\ddagger}$JD-163721 & 1.6145 & - & - & - & 1.35$\pm$0.19 & 8.88$\pm$0.03 & - & - \\
GT-169474 & 1.8404 & - & - & - & 0.46$\pm$0.07 & 8.65$\pm$0.03 & - & - \\
BA-173884 & 3.5328 & 6.89$\pm$0.63 & 3.13$\pm$0.31 & 1.81$\pm$0.19 & 0.34$\pm$0.02 & 8.59$\pm$0.01 & 0.90$\pm$0.18 & 2.63$\pm$0.44 \\
DD-149715 & 2.8107 & 7.22$\pm$0.59 & 3.30$\pm$0.28 & 1.28$\pm$0.13 & 0.43$\pm$0.02 & 8.64$\pm$0.01 & 1.04$\pm$0.09 & 2.41$\pm$0.16 \\
GT-173952 & 2.3039 & - & - & - & 0.24$\pm$0.04 & 8.52$\pm$0.03 & - & - \\
$^{\ddagger}$GT-175747 & 1.9168 & - & - & - & 1.23$\pm$0.39 & 8.86$\pm$0.07 & - & - \\
$^{\ddagger}$GR-172157 & 2.6935 & 5.68$\pm$5.56 & 2.43$\pm$1.18 & 1.05$\pm$1.24 & 0.53$\pm$0.07 & 8.69$\pm$0.03 & $<$1.67 & $>$0.07 \\
GT-162594 & 2.5264 & $>$0.89 & - & - & 0.63$\pm$0.14 & 8.72$\pm$0.05 & - & - \\
GR-175915 & 2.5750 & - & - & - & 1.10$\pm$0.13 & 8.84$\pm$0.02 & - & - \\
EX-172449 & 4.6229 & 2.55$\pm$0.13 & -0.30$\pm$0.17 & 3.32$\pm$0.17 & 0.19$\pm$0.01 & 8.46$\pm$0.01 & 1.06$\pm$0.26 & 2.37$\pm$0.82 \\
GR-146584 & 1.0372 & 18.10$\pm$8.50 & 6.46$\pm$1.05 & - & 0.49$\pm$0.02 & 8.67$\pm$0.01 & 0.77$\pm$0.11 & 2.89$\pm$0.26 \\
$^{\ddagger}$JD-172117 & 2.8233 & - & - & - & 0.79$\pm$0.09 & 8.77$\pm$0.02 & $<$1.51 & $>$0.07 \\
JD-176095 & 1.8070 & - & - & - & 0.46$\pm$0.16 & 8.66$\pm$0.08 & - & - \\
GR-172150 & 1.3172 & - & - & - & 1.22$\pm$0.09 & 8.86$\pm$0.02 & $<$0.92 & $>$2.59 \\
GR-172162 & 2.4542 & 5.13$\pm$2.06 & 2.14$\pm$0.98 & 1.06$\pm$6.32 & 0.49$\pm$0.06 & 8.67$\pm$0.03 & $<$1.27 & $>$1.90 \\
$^{\ddagger}$JD-146516 & 1.8469 & 5.03$\pm$0.81 & 2.04$\pm$0.53 & 0.88$\pm$0.20 & 0.40$\pm$0.04 & 8.63$\pm$0.02 & 0.95$\pm$0.17 & 2.56$\pm$0.38 \\
GR-172152 & 1.9981 & $>$2.93 & - & - & 0.79$\pm$0.13 & 8.77$\pm$0.03 & - & - \\
JD-172382 & 4.4374 & - & - & - & 0.13$\pm$0.01 & 8.39$\pm$0.01 & 1.02$\pm$0.14 & 2.43$\pm$0.27 \\
\hline
\end{tabular}
\tablefoot{
Observed (dust un-corrected) emission line properties for the 48 far-infrared detected NIRSpec sources, ranked in descending infrared luminosity, as per Figure \ref{fig:RGB}. For each source we list the NIRSpec ID (Column 1, C1),  NIRSpec spectroscopic redshift (C2), Balmer decrement (C3), [OIII]/H$\beta$ ratio (C5), \NH\ ratio (C6), \NH\ derived gas-phase metallicity (C7), [SII] doublet ratio (C8), [SII] derived electron density (C9). For low signal to noise emission lines detected we define a 3$\sigma$ limit in the respective calculation, as indicated by the inequalities. Sources identified as AGN (see Sec. \ref{Sec:AGN}) are highlighted, where $\dagger$ indicates broad emission lines, $\ddagger$ indicates an X-ray counterpart and $*$ highlights elevated optical-line ratios.
}
\end{table*}

\begin{table*}
\section{SED modelling}\label{App: SED_table}
\label{Table:SED}
\renewcommand{\arraystretch}{1}
\centering
\begin{tabular}{clccccccccc}
\hline
 ID & $z_{\rm spec}$ & log$_{10}$(M$_\ast$) &  SFR  & A$_{\rm V}$ & log$_{10}$(M$_{\rm d}$) & L$_{\rm FIR}$ & $T_{\rm d}$ & $R_{\rm e,0.5\mu m}$ & $n_{\rm 0.5\mu m}$ \\
 & & [M$_{\odot}$]& [M$_{\odot}$\,yr$^{-1}$] & [mag] &[M$_{\odot}$]& [L$_{\odot}$] & [K] & [kpc]  & \\
\hline
RB-160718 & 3.5444 & 10.69$\pm$0.08 & 67.04$\pm$41.81 & 2.76$\pm$0.46 & 9.06$^{+0.08}_{-0.09}$ & 12.69$^{+0.05}_{-0.06}$ & 37.37$^{+1.79}_{-1.74}$ & 1.36$^{+0.21}_{-0.21}$ & 0.67$^{+0.04}_{-0.02}$ \\
GT-167859 & 2.0836 & 10.98$\pm$0.04 & 64.51$\pm$23.53 & 0.70$\pm$0.14 & 9.18$^{+0.03}_{-0.03}$ & 12.63$^{+0.01}_{-0.02}$ & 34.67$^{+0.58}_{-0.56}$ & 4.04$^{+0.04}_{-0.04}$ & 0.93$^{+0.03}_{-0.03}$ \\
$^{\dagger\ddagger*}$GR-172113 & 2.5829 & 11.10$\pm$0.03 & 21.76$\pm$5.35 & 0.56$\pm$0.15 & 8.90$^{+0.02}_{-0.02}$ & 12.61$^{+0.01}_{-0.01}$ & 38.64$^{+0.31}_{-0.32}$ & 2.74$^{+0.01}_{-0.01}$ & 0.94$^{+0.01}_{-0.01}$ \\
$^{\ddagger}$JD-159764 & 2.5428 & 10.46$\pm$0.03 & 30.48$\pm$3.75 & 0.72$\pm$0.07 & 8.40$^{+0.09}_{-0.12}$ & 12.57$^{+0.03}_{-0.04}$ & 47.16$^{+2.03}_{-1.58}$ & 2.46$^{+0.06}_{-0.06}$ & 0.65$^{+0.00}_{-0.00}$ \\
$^{\dagger}$DD-160970 & 2.5448 & 11.14$\pm$0.10 & 112.85$\pm$60.00 & 3.01$\pm$0.48 & 9.40$^{+0.04}_{-0.04}$ & 12.48$^{+0.05}_{-0.06}$ & 30.24$^{+1.01}_{-1.01}$ & 5.25$^{+0.29}_{-0.25}$ & 1.12$^{+0.09}_{-0.09}$ \\
$^{\dagger}$BA-173854 & 2.5554 & 11.41$\pm$0.06 & 189.09$\pm$108.81 & 3.09$\pm$0.63 & 9.43$^{+0.05}_{-0.05}$ & 12.47$^{+0.02}_{-0.02}$ & 29.79$^{+0.70}_{-0.66}$ & 3.51$^{+0.13}_{-0.12}$ & 1.92$^{+0.09}_{-0.08}$ \\
$^{\ddagger}$JD-172437 & 2.3032 & 11.41$\pm$0.07 & 98.05$\pm$47.01 & 3.81$\pm$0.23 & 8.89$^{+0.05}_{-0.06}$ & 12.41$^{+0.02}_{-0.02}$ & 35.62$^{+0.68}_{-0.60}$ & - & - \\
DD-147398 & 2.9436 & 11.29$\pm$0.09 & 0.13$\pm$103.03 & 2.31$\pm$0.77 & 9.18$^{+0.08}_{-0.08}$ & 12.40$^{+0.05}_{-0.05}$ & 31.92$^{+1.33}_{-1.24}$ & - & - \\
GR-172160 & 2.6974 & 10.90$\pm$0.03 & 60.36$\pm$11.93 & 2.53$\pm$0.17 & 8.84$^{+0.10}_{-0.12}$ & 12.36$^{+0.04}_{-0.05}$ & 35.72$^{+1.20}_{-1.02}$ & 3.61$^{+0.05}_{-0.05}$ & 0.92$^{+0.03}_{-0.03}$ \\
GT-162506 & 2.1562 & 11.49$\pm$0.07 & 303.11$\pm$90.16 & 2.10$\pm$0.14 & 9.17$^{+0.07}_{-0.07}$ & 12.35$^{+0.06}_{-0.06}$ & 31.36$^{+1.31}_{-1.27}$ & 5.53$^{+0.27}_{-0.23}$ & 0.86$^{+0.08}_{-0.07}$ \\
GT-162118 & 1.7059 & 11.06$\pm$0.07 & 76.95$\pm$28.09 & 2.71$\pm$0.24 & 9.24$^{+0.06}_{-0.06}$ & 12.33$^{+0.04}_{-0.05}$ & 30.36$^{+1.03}_{-0.99}$ & 3.98$^{+0.04}_{-0.04}$ & 0.87$^{+0.02}_{-0.02}$ \\
JD-173408 & 4.4319 & 10.69$\pm$0.03 & 56.19$\pm$8.01 & 1.68$\pm$0.08 & 9.14$^{+0.09}_{-0.12}$ & 12.20$^{+0.09}_{-0.12}$ & 30 & 1.99$^{+0.02}_{-0.02}$ & 0.93$^{+0.03}_{-0.03}$ \\
$^{\dagger\ddagger*}$GR-172166 & 2.2245 & 10.91$\pm$0.04 & 263.68$\pm$41.87 & 1.44$\pm$0.06 & 8.62$^{+0.09}_{-0.10}$ & 12.20$^{+0.03}_{-0.03}$ & 36.57$^{+1.25}_{-1.02}$ & 2.09$^{+0.01}_{-0.01}$ & 3.99$^{+0.03}_{-0.03}$ \\
$^{\dagger\ddagger*}$JD-149659 & 2.6138 & 10.50$\pm$0.05 & 3.81$\pm$6.07 & 1.56$\pm$0.45 & 7.75$^{+0.18}_{-0.23}$ & 12.19$^{+0.05}_{-0.06}$ & 53.72$^{+5.48}_{-3.67}$ & 1.42$^{+0.01}_{-0.01}$ & 0.66$^{+0.01}_{-0.01}$ \\
RB-151060 & 4.5440 & 10.66$\pm$0.08 & 64.71$\pm$27.40 & 3.36$\pm$0.35 & 9.09$^{+0.04}_{-0.04}$ & 12.15$^{+0.04}_{-0.04}$ & 30 & 2.02$^{+0.33}_{-0.26}$ & 1.01$^{+0.31}_{-0.24}$ \\
JD-170092 & 3.7476 & 10.83$\pm$0.06 & 55.68$\pm$36.42 & 3.61$\pm$0.60 & 9.05$^{+0.06}_{-0.06}$ & 12.11$^{+0.06}_{-0.06}$ & 30 & 12.14$^{+3.14}_{-2.75}$ & 1.61$^{+0.46}_{-0.38}$ \\
$^{\ddagger}$JD-167963 & 2.0346 & 11.16$\pm$0.03 & 64.30$\pm$9.91 & 3.94$\pm$0.08 & 8.74$^{+0.10}_{-0.11}$ & 12.09$^{+0.03}_{-0.03}$ & 33.45$^{+1.20}_{-1.08}$ & 2.48$^{+0.01}_{-0.01}$ & 1.02$^{+0.01}_{-0.01}$ \\
RB-161829 & 3.9706 & 10.79$\pm$0.08 & 141.57$\pm$80.83 & 2.16$\pm$0.22 & 9.02$^{+0.08}_{-0.09}$ & 12.08$^{+0.08}_{-0.09}$ & 30 & 2.31$^{+0.06}_{-0.06}$ & 0.78$^{+0.06}_{-0.07}$ \\
$^{\dagger\ddagger}$JD-176494 & 2.5834 & 10.71$\pm$0.04 & 11.48$\pm$5.64 & 1.11$\pm$0.19 & 7.74$^{+0.26}_{-0.29}$ & 12.07$^{+0.21}_{-0.37}$ & 51.18$^{+13.40}_{-11.81}$ & 1.19$^{+0.01}_{-0.01}$ & 1.40$^{+0.03}_{-0.04}$ \\
GT-169541 & 2.3032 & 10.96$\pm$0.06 & 56.71$\pm$17.71 & 3.76$\pm$0.33 & 8.39$^{+0.14}_{-0.15}$ & 12.07$^{+0.05}_{-0.05}$ & 38.08$^{+2.70}_{-2.35}$ & 4.41$^{+0.50}_{-0.40}$ & 2.99$^{+0.31}_{-0.29}$ \\
$^{\ddagger}$JD-164819 & 1.6153 & 10.58$\pm$0.08 & 12.28$\pm$19.48 & 1.15$\pm$0.46 & 8.88$^{+0.09}_{-0.11}$ & 12.03$^{+0.02}_{-0.02}$ & 31.09$^{+1.11}_{-0.92}$ & - & - \\
$^{\dagger}$GR-175918 & 2.3279 & 10.72$\pm$0.06 & 0.00$\pm$6.29 & 1.16$\pm$0.36 & 8.41$^{+0.19}_{-0.24}$ & 12.00$^{+0.06}_{-0.07}$ & 36.66$^{+2.91}_{-1.94}$ & 1.76$^{+0.04}_{-0.04}$ & 3.78$^{+0.11}_{-0.10}$ \\
RB-163266 & 4.8041 & 10.73$\pm$0.08 & 93.63$\pm$41.58 & 2.34$\pm$0.24 & 8.91$^{+0.10}_{-0.12}$ & 11.96$^{+0.10}_{-0.12}$ & 30 & 1.62$^{+0.99}_{-0.43}$ & 3.40$^{+1.82}_{-1.13}$ \\
$^{\dagger\ddagger}$GT-162424 & 4.5560 & 11.47$\pm$0.01 & 0.00$\pm$0.10 & 0.01$\pm$0.01 & 8.91$^{+0.05}_{-0.06}$ & 11.96$^{+0.05}_{-0.06}$ & 30 & 0.14$^{+0.01}_{-0.01}$ & 1.59$^{+1.12}_{-0.90}$ \\
BA-155180 & 2.0977 & 11.08$\pm$0.07 & 85.73$\pm$33.11 & 2.97$\pm$0.23 & 8.88$^{+0.08}_{-0.10}$ & 11.93$^{+0.08}_{-0.10}$ & 30 & 4.31$^{+0.08}_{-0.09}$ & 0.72$^{+0.03}_{-0.03}$ \\
GR-172171 & 1.8130 & 10.67$\pm$0.04 & 3.83$\pm$16.31 & 2.15$\pm$0.33 & 8.85$^{+0.02}_{-0.02}$ & 11.91$^{+0.02}_{-0.02}$ & 30 & 5.80$^{+0.03}_{-0.03}$ & 0.83$^{+0.01}_{-0.01}$ \\
RB-157952 & 5.2172 & 10.81$\pm$0.12 & 190.14$\pm$66.97 & 2.97$\pm$0.24 & 8.83$^{+0.08}_{-0.09}$ & 11.89$^{+0.08}_{-0.09}$ & 30 & 3.37$^{+0.64}_{-0.47}$ & 2.11$^{+0.53}_{-0.42}$ \\
JD-175480 & 2.5651 & 10.57$\pm$0.06 & 30.04$\pm$10.77 & 2.25$\pm$0.29 & 8.78$^{+0.05}_{-0.06}$ & 11.83$^{+0.05}_{-0.06}$ & 30 & - & - \\
$^{\dagger}$BA-173819 & 2.4934 & 10.34$\pm$0.03 & 15.74$\pm$1.43 & 0.17$\pm$0.09 & 6.90$^{+0.60}_{-0.53}$ & 11.83$^{+0.10}_{-0.16}$ & 71.86$^{+29.42}_{-21.23}$ & 3.25$^{+0.13}_{-0.13}$ & 2.13$^{+0.17}_{-0.17}$ \\
DD-150418 & 1.8187 & 11.34$\pm$0.07 & 108.56$\pm$54.74 & 3.40$\pm$0.35 & 8.75$^{+0.08}_{-0.09}$ & 11.81$^{+0.08}_{-0.09}$ & 30 & 4.28$^{+0.05}_{-0.06}$ & 0.71$^{+0.03}_{-0.03}$ \\
$^{\ddagger}$JD-163721 & 1.6145 & 11.26$\pm$0.06 & 46.72$\pm$22.89 & 2.50$\pm$0.28 & 9.35$^{+0.09}_{-0.10}$ & 11.81$^{+0.03}_{-0.03}$ & 24.49$^{+0.76}_{-0.64}$ & 4.90$^{+0.02}_{-0.02}$ & 2.20$^{+0.02}_{-0.02}$ \\
GT-169474 & 1.8404 & 11.14$\pm$0.08 & 123.55$\pm$62.95 & 3.13$\pm$0.23 & 8.42$^{+0.66}_{-0.34}$ & 11.71$^{+0.15}_{-1.36}$ & 33.26$^{+5.57}_{-16.00}$ & 3.11$^{+0.03}_{-0.03}$ & 0.95$^{+0.02}_{-0.02}$ \\
BA-173884 & 3.5328 & 10.71$\pm$0.08 & 152.29$\pm$45.56 & 1.73$\pm$0.14 & 8.64$^{+0.05}_{-0.06}$ & 11.70$^{+0.05}_{-0.06}$ & 30 & 2.49$^{+0.08}_{-0.07}$ & 1.12$^{+0.08}_{-0.07}$ \\
DD-149715 & 2.8107 & 10.89$\pm$0.08 & 143.63$\pm$44.18 & 1.99$\pm$0.17 & 8.64$^{+0.03}_{-0.03}$ & 11.69$^{+0.03}_{-0.03}$ & 30 & 2.81$^{+0.08}_{-0.08}$ & 1.39$^{+0.08}_{-0.09}$ \\
GT-173952 & 2.3039 & 10.94$\pm$0.06 & 67.59$\pm$24.51 & 3.10$\pm$0.27 & 8.63$^{+0.05}_{-0.06}$ & 11.69$^{+0.05}_{-0.06}$ & 30 & 4.44$^{+0.37}_{-0.28}$ & 1.56$^{+0.18}_{-0.15}$ \\
$^{\ddagger}$GT-175747 & 1.9168 & 10.98$\pm$0.06 & 59.11$\pm$18.72 & 2.72$\pm$0.20 & 8.54$^{+0.06}_{-0.07}$ & 11.59$^{+0.06}_{-0.07}$ & 30 & 2.59$^{+0.04}_{-0.04}$ & 0.94$^{+0.05}_{-0.05}$ \\
$^{\ddagger}$GR-172157 & 2.6935 & 10.73$\pm$0.04 & 43.55$\pm$9.00 & 2.20$\pm$0.10 & 8.51$^{+0.12}_{-0.17}$ & 11.56$^{+0.12}_{-0.17}$ & 30 & 4.99$^{+0.10}_{-0.09}$ & 1.78$^{+0.04}_{-0.04}$ \\
GT-162594 & 2.5264 & 10.60$\pm$0.07 & 39.12$\pm$18.91 & 2.11$\pm$0.33 & 8.49$^{+0.07}_{-0.09}$ & 11.55$^{+0.07}_{-0.09}$ & 30 & 3.34$^{+0.16}_{-0.14}$ & 0.72$^{+0.06}_{-0.05}$ \\
GR-175915 & 2.5750 & 10.08$\pm$0.05 & 0.01$\pm$1.33 & 1.05$\pm$0.44 & 8.45$^{+0.10}_{-0.13}$ & 11.51$^{+0.10}_{-0.13}$ & 30 & 1.65$^{+0.04}_{-0.04}$ & 0.70$^{+0.05}_{-0.03}$ \\
EX-172449 & 4.6229 & 9.96$\pm$0.08 & 29.04$\pm$6.76 & 0.48$\pm$0.11 & 8.42$^{+0.04}_{-0.04}$ & 11.47$^{+0.04}_{-0.04}$ & 30 & 1.36$^{+0.04}_{-0.04}$ & 1.57$^{+0.11}_{-0.12}$ \\
GR-146584 & 1.0372 & 10.78$\pm$0.06 & 0.43$\pm$1.76 & 2.26$\pm$0.48 & 8.07$^{+0.05}_{-0.05}$ & 11.47$^{+0.01}_{-0.01}$ & 34.09$^{+0.60}_{-0.61}$ & 2.06$^{+0.01}_{-0.01}$ & 1.04$^{+0.01}_{-0.01}$ \\
$^{\ddagger}$JD-172117 & 2.8233 & 10.99$\pm$0.04 & 0.00$\pm$0.20 & 2.32$\pm$0.36 & 8.35$^{+0.16}_{-0.31}$ & 11.41$^{+0.16}_{-0.31}$ & 30 & 2.18$^{+0.03}_{-0.03}$ & 2.24$^{+0.09}_{-0.07}$ \\
JD-176095 & 1.8070 & 10.51$\pm$0.07 & 15.82$\pm$9.85 & 1.42$\pm$0.31 & 8.30$^{+0.12}_{-0.17}$ & 11.36$^{+0.12}_{-0.17}$ & 30 & - & - \\
GR-172150 & 1.3172 & 10.89$\pm$0.07 & 18.54$\pm$10.84 & 2.37$\pm$0.34 & 7.66$^{+0.18}_{-0.17}$ & 11.29$^{+0.05}_{-0.05}$ & 37.35$^{+2.23}_{-2.15}$ & 3.23$^{+0.01}_{-0.01}$ & 2.48$^{+0.03}_{-0.03}$ \\
GR-172162 & 2.4542 & 10.51$\pm$0.07 & 39.34$\pm$13.87 & 2.41$\pm$0.20 & 8.07$^{+0.07}_{-0.09}$ & 11.12$^{+0.07}_{-0.09}$ & 30 & 2.64$^{+0.05}_{-0.05}$ & 1.14$^{+0.03}_{-0.03}$ \\
$^{\ddagger}$JD-146516 & 1.8469 & 10.39$\pm$0.03 & 104.24$\pm$37.03 & 1.44$\pm$0.07 & 8.06$^{+0.07}_{-0.09}$ & 11.11$^{+0.07}_{-0.09}$ & 30 & 4.08$^{+0.01}_{-0.01}$ & 0.92$^{+0.01}_{-0.01}$ \\
GR-172152 & 1.9981 & 10.24$\pm$0.06 & 0.65$\pm$1.93 & 1.68$\pm$0.27 & 7.99$^{+0.08}_{-0.11}$ & 11.05$^{+0.08}_{-0.11}$ & 30 & 2.14$^{+0.01}_{-0.01}$ & 1.24$^{+0.02}_{-0.02}$ \\
JD-172382 & 4.4374 & 9.68$\pm$0.09 & 57.26$\pm$12.73 & 1.06$\pm$0.05 & 7.89$^{+0.13}_{-0.20}$ & 10.94$^{+0.13}_{-0.20}$ & 30 & 1.63$^{+0.04}_{-0.04}$ & 0.90$^{+0.05}_{-0.06}$ \\
\hline
\end{tabular}
\tablefoot{
The SED properties for the 48 far-infrared detected NIRSpec sources, ranked in descending infrared luminosity, as per Figure \ref{fig:RGB}. For each source we list the \tcr{NIRSpec} ID (Column 1, C1), NIRSpec spectroscopic redshift (C2), \texttt{bagpipes} derived stellar mass (C3), star-formation rate (C4) and dust attenuation (C5). We further list the \texttt{mercurius} derived dust mass (C6), far-infrared luminosity (C7), dust temperature (C8), as well as the \texttt{PySersic} derived 0.5\um\ effective radius (C9) and S\'ersic index (C10). Sources identified as AGN (see Sec. \ref{Sec:AGN}) are highlighted, where $\dagger$ indicates broad emission lines, $\ddagger$ indicates an X-ray counterpart and $*$ highlights elevated optical-line ratios.
}
\end{table*}

\end{appendix}
\end{document}